\documentclass[12pt,a4paper,
preprint]{iopart}

\usepackage{iopams}  
\usepackage{graphicx}
\usepackage[breaklinks=true,colorlinks=true,linkcolor=blue,urlcolor=blue,citecolor=blue]{hyperref}
\eqnobysec

\begin{document}

\renewcommand{\L}{{\cal L}}
\renewcommand{\l}{\mathfrak l}

\begin{flushright}
UTHEP-681 \\
RIKEN-TH-210 \\
RIKEN-QHP-211\\
RIKEN-STAMP-24\\
January 2016\\
\end{flushright}
\title[Dipolar quantization and 2dCFT]{Dipolar quantization and the infinite circumference limit of two-dimensional conformal field theories}

\author{Nobuyuki  Ishibashi
}
\address{
Faculty of Pure and Applied Sciences, University of Tsukuba, \\
Tsukuba, Ibaraki 305-8571, Japan }
\ead{ishibash@het.ph.tsukuba.ac.jp}

\author[cor1]{Tsukasa Tada}
\address{RIKEN Nishina Center for Accelerator-based Science, \\
Wako, Saitama 351-0198, Japan}
\ead{\mailto{tada@riken.jp}}

\begin{abstract}
Elaborating on our previous presentation, where the term {\it dipolar quantization} was introduced, we argue here that adopting $L_0-(L_1+L_{-1})/2+{\bar L}_0-({\bar L}_1+{\bar L}_{-1})/2$ as the Hamiltonian instead of $L_0+{\bar L}_0$ yields an infinite circumference limit  in two-dimensional conformal field theory. The new Hamiltonian leads to dipolar quantization instead of  radial quantization. As a result, the new theory exhibits a continuous and strongly degenerated spectrum in addition to the Virasoro algebra with a continuous index. Its Hilbert space exhibits a different inner product than that obtained in the original theory. The idiosyncrasy of this particular Hamiltonian is its relation to the so-called sine-square deformation, which is found in the study of a certain class of quantum statistical systems. The appearance of the infinite circumference explains why the vacuum states  of sine-square deformed systems are coincident with those of the respective closed-boundary systems.
\end{abstract}


\section{Introduction}
Quantum field theory represents undoubtedly one of the greatest pinnacles of human knowledge. Although its profoundness and versatility have nurtured a number of novel and important concepts along its nearly a century-long history \cite{Heisenberg:1929xj},  it continues to reveal marvellous features and to produce many new findings. In many of these discoveries, the idea of symmetry has played an essential and exceptional role. In particular, focusing on the proper symmetry and delving into its meaning has been proven to be the most fruitful strategy.
A recent example is the role played by $SO(2,4)$ symmetry in AdS/CFT correspondence \cite{Maldacena:1997re}.

In this report, we offer another  example of quantum field theory where symmetry plays a central role and reveals interesting phenomena, expanding on our preceding presentation \cite{Ishibashi:2015jba}. Our focal point here is the global conformal symmetry  
in two-dimensional conformal field theory (2d CFT), which is homomorphic to $SL(2,\mathbb{R})$.
\footnote{See \cite{Miyaji:2015fia,Verlinde:2015qfa,Nakayama:2015mva} for recent attempts to exploit $SL(2,\mathbb{R})$ symmetry in the context of AdS$_{3}$/CFT$_{2}$ correspondence.
Ref. \cite{Strominger:2013lka} and the references there in also offer enthralling perspectives on the role played by $SL(2,\mathbb{C}) \sim SL(2,\mathbb{R})\times SL(2,\mathbb{R})$.
}

The Virasoro algebra, which is an infinite dimensional Lie algebra, dictates the symmetry of 2d CFT. A notable subalgebra of the Virasoro algebra is the one that generates the global conformal transformation on the two-dimensional worldsheet where the CFT resides. The transformation consists of  $L_0$, $L_1$ and $L_{-1}$ generators, and their anti-holomorphic counterparts, which we omit for the sake of simplicity. Then, it turns out to be isomorphic to $sl(2,\mathbb{R})$.

\begin{figure}[tbh]
\centering
\includegraphics[width=7cm]{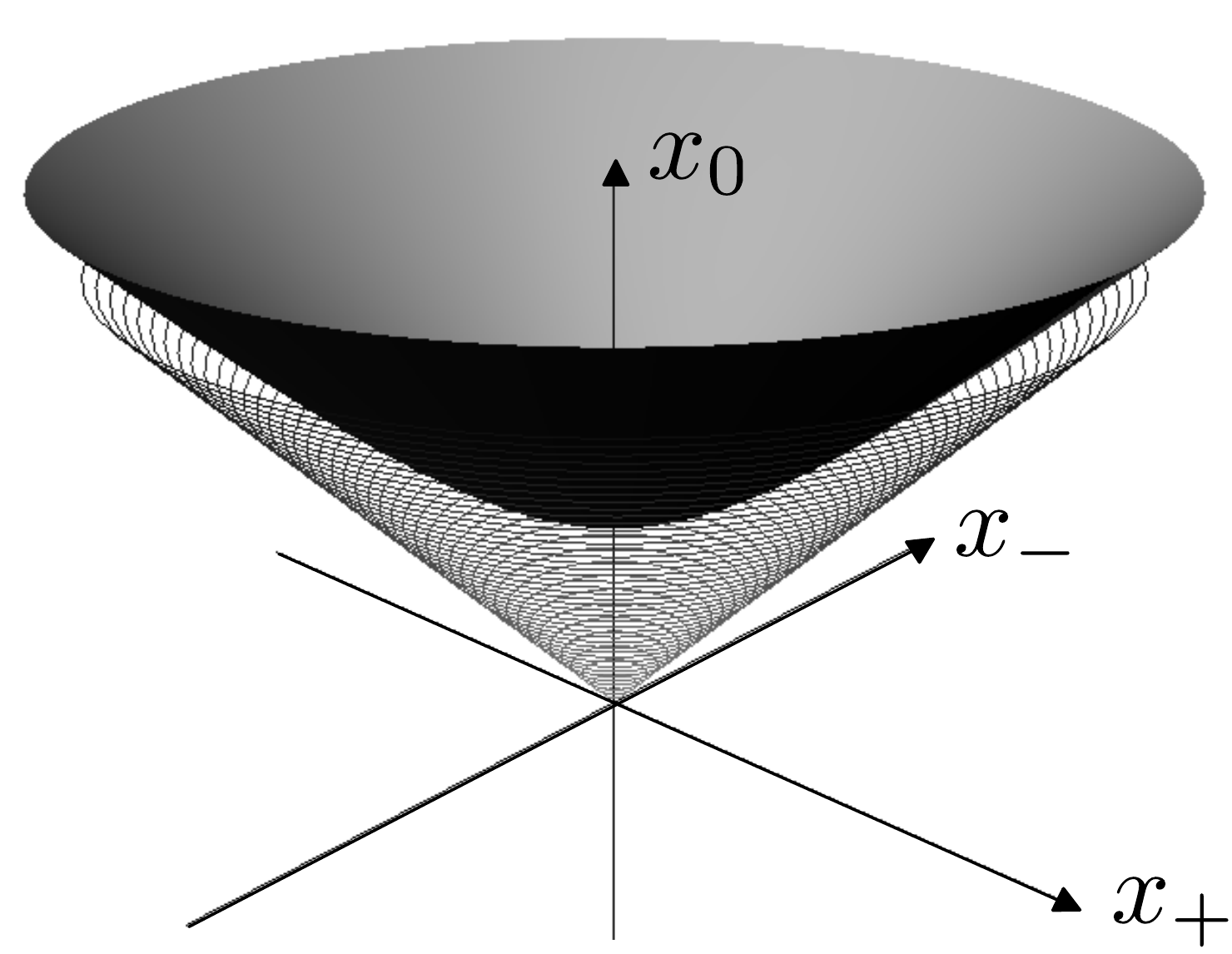}
\caption{The upper half of a hyperboloid and a surrounding (light-) cone are rendered  in the $(x_{0},x_{+},x_{-})$ space. Any point on the hyperboloid is $SL(2, \mathbb{R})$-equivalent to an arbitrary point on the hyperboloid, particularly to the lowest point on the hyperboloid.}
\label{fig:hyperboloid}
\end{figure}

It would now be helpful to establish the relationship between $sl(2,\mathbb{R})$ and the aforementioned subalgebra of the Virasoro algebra, by introducing the following new linear combination of the generators:
\begin{equation}
L_+=\frac{L_1+L_{-1}}{2} \ ,\ L_-=\frac{L_1-L_{-1}}{2i}.
\end{equation}
The Casimir operator of the subalgebra can be expressed in a more familiar manner:
\begin{equation}
C_2=L_0^2-L_+^2-L_-^2.
\end{equation}
The global conformal transformation can be represented naturally as the adjoint action over the space spanned by $L_{0}, L_{+}$ and $L_{-}$:
\begin{equation}
x_{0}L_{0}+x_{+}L_{+}+x_{-}L_{-}.
\end{equation}
The adjoint action alters the coefficients $x_{0}, x_{+}$ and  $x_{-}$, 
while retaining
\begin{equation}
x_{0}^{2}-x_{+}^{2}-x_{-}^{2} \label{eqn:proper0}
\end{equation}
invariant, and it generates the corresponding change in the worldsheet coordinates due to the global conformal transformation. Different sets of $x_{0}, x_{+}$ and $x_{-}$ are connected through the global conformal transformation or $SL(2,\mathbb{R})$. For example, any point on the hyperboloid depicted in Fig. \ref{fig:hyperboloid} can be  converted to  the bottom of the hyperboloid, $(x_{0}, x_{+}, x_{-})=(1,0,0)$ with an appropriate transformation, as is clear from the invariance of the expression (\ref{eqn:proper0}). The significance of the point represented by $(1,0,0)$ is its correspondence with the generator $L_{0}$, which is the Hamiltonian of the theory.
Thus, the Hamiltonian $L_{0}$ is stable against  perturbations caused by adding a small amount of $L_{+}$ or $L_{-}$, in the sense that small $x_{+}$ and $x_{-}$ coefficients can be annihilated by the $SL(2, \mathbb{R})$ action, as explained above, and then $x_{0}$ can also be rescaled to unity if desired.

From this perspective on $SL(2, \mathbb{R})$ symmetry, Fig. \ref{fig:hyperboloid} reveals that in the parameter space of $(x_{0},x_{+},x_{-})$, there exist two other disconnected regions that should have their own physical significance.  The first is the meshed cone to which the hyperboloid asymptotes, and to which we refer  hereafter as the `light-cone' in comparison with the three-dimensional (3d) Lorentz geometry. The light-cone is represented by the point $(x_{0},x_{+},x_{-})=(1,-1,0)$, or  the generator $L_{0}-L_{+}$. Outside of the light-cone is the second region where the other type of hyperboloids can be placed. The second region is represented by $(x_{0},x_{+},x_{-})=(0,1,0)$, or  $L_{+}=(L_{1}+L_{-1})/2$. Since the representative generator $L_{0}$ plays a central role in CFT as a Hamiltonian,  the representatives of the other two distinct regions, $L_{0}-L_{+}$ and $L_{+}$, may play important roles as well.

From a symmetry viewpoint, the Hilbert space of a quantum system is the representation vector space of the symmetry. To construct the entire representation vector space of a large symmetry algebra, it is often useful to first investigate its smaller subalgebra. Once we know the subalgebra's irreducible representations, which can be classified by the value of the Casimir operator, the representation space for the whole symmetry algebra can be constructed from the direct sum of these irreducible representations (of the subalgebra). Although the action of the subalgebra remains within each irreducible representations, the rest of the generators of the entire symmetry algebra are generally represented by the transitions between the various irreducible representation sectors of the subalgebra. In this way, the structure of the representation of the subalgebra is reflected into the Hilbert space of the quantum system. Thus, the considerations in the previous paragraph  should certainly shed light on the structure of the Hilbert space such as the spectrum of CFT.

In fact, the spectrum of $L_{0}$ fits nicely in the discrete representation of $sl(2, \mathbb{R})$. Although the lack of translational symmetry renders the analogy with the 3d Poincar\'e representation  imperfect, it is natural to note it as a `massive' spectrum. In this case, another region including $L_{0}-L_{+}$ could be called a `massless' representation. Because the mass scale stems from the size of the system, or the finite scale of a CFT \cite{Belavin:1984vu,Cardy:1984rp}, space is implied to be infinite in size. In addition, the continuous nature of the massless representation suggests the emergence of a continuous index of the symmetry algebra. As a matter of fact,  continuous representations of $sl(2, \mathbb{R})$ do exist \cite{Wybourne:1974}, and we will uncover the continuous Virasoro algebra in this report. The situation here may be compared with the light-front quantization \cite{Dirac:1949cp}, or the infinite momentum frame \cite{Weinberg:1966jm}.

Changing the Hamiltonian also affects the structure of the Hilbert space; hence, we will be dealing with an entirely different Hilbert space.
The most graphical change would be the feature of the time-translational vector, which was radial for  time translation generated by $L_{0}$. The time-translational fields radiate from the origin just as  electrical flux does from a positive charge. These fluxes then converge to a (fictitious) negative charge at  infinity. The origin and infinity correspond to $t\rightarrow -\infty$ and $t\rightarrow \infty$, respectively.
On the other hand, $L_{0}-L_{+}$ produces a vector field that is similar to the dipolar electric field. We thus named the present procedure  `dipolar quantization'. In this case, the infinite past $t\rightarrow-\infty$ and the infinite future $t\rightarrow \infty$ are both located at $z=1$ with the (supposedly) infinitesimal separation, just as a dipole consists of the negative and positive charges at the same point. Therefore, in analogy with the terminology of `radial quantization', we name the present procedure dipolar quantization \cite{Ishibashi:2015jba}.
While  conformal symmetry dictates the behaviour of $T(z)$ and the primary operators $\phi(z)$ in the $z$ coordinate, a different time translation or a different equal-time contour affects the Hilbert space structure. This effect stems from the salient connection between spacetime and the operators in quantum field theories.

The reader might be curious how the analysis presented here relates to the study of tensionless strings. Although tensionless strings naturally tend to spread freely, the symmetry that tensionless strings exhibit is a (2d) Galilean conformal algebra, which differs from the continuous Virasoro algebra that we find here. For recent research on tensionless strings, see  \cite{Bagchi:2015nca} and the references therein\footnote{As a matter of fact, the relationship between tensionless closed and open strings was  discussed in \cite{Bagchi:2015nca}, and so we also seem to share a common motivation.}.

The present research is partially motivated by the phenomenon known as the sine-square deformation (SSD) \cite{SSD}. Reference \cite{SSD} and subsequent studies \cite{Proof1, Proof2,ShibataHotta2011,HikiharaNishinio:2011,Gendiaretal2011} revealed that (i) if we consider the vacuum of a certain class of (one-dimensional) quantum systems of size $L$ with the closed boundary condition and (ii) construct the vacuum of the same system with the open boundary condition and an additional space-dependent ($x$-dependent)  modulation of the coupling constant as
\begin{equation}
g\sin^{2}(2\pi\frac{x}{L}), \label{eqn:ssd1}
\end{equation}
then these two vacua are identical. Note that in Eq.(\ref{eqn:ssd1}), the coupling constant goes to
zero at the both ends of the system, which leads to the open boundary.

If SSD were employed for string theory, this coincidence implies that the closed string vacuum is identical to the open string vacuum in a certain (worldsheet) `background'. In fact, it was shown that SSD is also applicable to CFTs \cite{Katsura:2011ss}. When applied to CFT, the modulation (\ref{eqn:ssd1}) amounts to changing the Hamiltonian to
\begin{equation}
L_{0}-\frac{L_{1}+L_{-1}}{2} +{\bar L}_{0}-\frac{{\bar L}_{1}+{\bar L}_{-1}}{2} .
\end{equation}
The holomorphic part of the above is nothing but $L_{0}-L_{+}$, which we  introduced earlier.
The implications of SSD to string theory were discussed in Refs. \cite{Tada:2014kza,Tada:2014jps}.

To date, studies of SSD have been limited to  studying the ground states with apparently different boundary conditions. However, the ground state contains essential information on the quantum system. For example, one can read off the central charge of a CFT from the ground-state entanglement \cite{Holzhey:1994we, Callan:1994py, Calabrese:2004eu} (see also Ref.\cite{Garrison:2015lva} for recent work). Therefore, one might expect that the significance of SSD extends beyond the ground states to the excited states. By clarifying the structure of the excited states under SSD, this report justifies this expectation

Conversely, it is mystifying that  systems with different boundary conditions have a common vacuum state.  Intuitively, the lowest-energy state should be the most vulnerable to the  influence of the global structure of the system, such as the boundary conditions. To clarify this conundrum, note that SSD systems possess continuum spectra, which implies that such systems have an infinitely large space. Then, the distinction between the open- and closed-conditions at the ends that located infinitely far away becomes irrelevant. It would be interesting to see if this explanation applies to other SSD systems besides the CFT systems considered here. In fact, Ref. \cite{Hotta:2014} suggests that there are  other systems that also show continuous spectra. 

The composition of the paper is as follows: Section \ref{sec:GA} is devoted to geometrical analysis, where we determine the differential operator that corresponds to the time translation driven by the new Hamiltonian. By using this differential operator, we establish the equal-time contour for this time translation. Section \ref{sec:QA}, investigates the quantum aspects of the system based on the geometric knowledge discussed in Section \ref{sec:GA}. Comprehension of the equal-time contour enables us to define the conserved charges. By calculating the commutation relations between these charges, we obtain the continuous Virasoro algebra and, in particular, the central extension term.  In Sec. \ref{sec:HCHS}, we further discuss the structure of the Hilbert space, mainly from the viewpoint of Hermitian conjugation .
We conclude and summarise the results in Sec. \ref{sec:SMDI} and raise several points worth discussing. We also supplement the main text with three appendices. In \ref{sec:HTDO}, we assemble various formulas in regard to the relevant differential operators. A keen reader might have noticed that we could have chosen a set $\{L_n, L_0, L_{-n} \}$ as the generators of the $SL(2, \mathbb{R})$ subalgebra of the Virasoro algebra: this case is explored in \ref{sec:ZnETC}. Our investigation here is restricted to CFT that resides strictly on a Riemann sphere which possess a unique spin structure. Although this fact is not explicitly exploited in the main text,  we  describe how the present formalism can be applied to superconformal field theories (SCFT) in  \ref{sec:SCFT}, where only one fermionic mode appears.

\section{Geometrical analysis}\label{sec:GA}
\subsection{The Witt algebra}

To clarify our notation, we briefly recapitulate the standard argument of the conformal transformation in two dimensions. For any field theory,  a constant translation
\begin{equation}
x_{\nu} \rightarrow x'_{\nu} =x_{\nu}+\epsilon_{\nu},
\end{equation}
evokes the associated Noether current that is called the energy-momentum tensor
\begin{equation}
J_{\mu}^{(\nu)}(x)=T_{\mu\nu}(x),
\end{equation}
where the extra index $\nu$ indicates the direction of the translation. Assuming invariance under translation, the Noether current $J_{\mu}^{(\nu)}$ is,  subject to the equation of conservation:
\begin{equation}
\partial^{\mu}J_{\mu}^{(\nu)}=\partial^{\mu}T_{\mu\nu}=0.
\end{equation}
Furthermore,  the energy-momentum tensor is known to be symmetric and traceless for conformal field theories:
\begin{equation}
T_{\mu\nu}=T_{\nu\mu}\ , \ T^{\mu}_{\mu}=0.
\end{equation}

The (infinitesimal) conformal transformation is the space-dependent (infinitesimal) translation
\begin{equation}
x_{\mu} \rightarrow x'_{\mu} =x_{\mu}+\epsilon_{\mu}(x)\,
\end{equation}
that keeps conformal flatness
\begin{equation}
g_{\mu\nu}(x)  \rightarrow e^{2\omega(x)} g_{\mu\nu}(x).
\end{equation}
This is achieved if the following conformal Killing equation is satisfied:
\begin{equation}
\nabla_{\mu}\epsilon_{\nu}+\nabla_{\nu}\epsilon_{\mu}=2\omega(x)g_{\mu\nu}(x).\label{eqn:conflkill}
\end{equation}
A vector field $\epsilon_{\mu}$ that satisfies the conformal Killing equation (\ref{eqn:conflkill}) is called the conformal Killing vector.
Conserved currents associated with the conformal transformation can be constructed by using the conformal Killing vector $\epsilon_{\mu}$ and the energy-momentum tensor as follows:
\begin{equation}
J_{\mu}^{\epsilon}(x)=\epsilon^{\nu}(x) T_{\mu\nu}(x).
\end{equation}

If spacetime is  2d Euclidean, the conformal Killing equation reads
\begin{equation}
\partial_{a} \epsilon_{b} + \partial_{b} \epsilon_{a}= 2\omega(x) \delta_{ab}\  , \quad a, b=0,1 .
\end{equation}
The above equation retains the same structure as the Cauchy-Riemann relation; hence, the conformal transformation in two dimensions can be expressed in terms of the analytic coordinate transformation
\begin{equation}
z \rightarrow \epsilon(z) \  , \  {\bar z} \rightarrow {\bar \epsilon}({\bar z}), \label{eqn:zezbeb}
\end{equation}
if we adopt the complex coordinate system $z=x^{1}+ix^{2}$. The conserved (Noether) current can also be expressed in terms of the complex coordinate as
\begin{equation}
J_{0}^{\epsilon}=T_{zz}(z)\epsilon(z)+{\bar T}_{{\bar z}{\bar z}}({\bar z}){\bar \epsilon}({\bar z}) \ , \ 
J_{1}^{\epsilon}=i\left( T_{zz}(z)\epsilon(z)
-
{\bar T}_{{\bar z}{\bar z}}({\bar z}){\bar \epsilon}({\bar z})\right). \label{eqn:JTe}
\end{equation}

In particular, upon introducing the specific set of infinitesimal transformations
\begin{equation}
z \rightarrow z' =z-\epsilon_{n} z^{n+1}\ , \ {\bar z} \rightarrow {\bar z}' ={\bar z}-{\bar \epsilon}_{n} {\bar z }^{n+1}, \label{eqn:enn}
\end{equation}
the generators for the transformations at issue can be written as
\begin{equation}
l_{n}=-z^{n+1} \frac{\partial}{\partial z}\ , \ {\bar l}_{n}=-{\bar z}^{n+1} \frac{\partial}{\partial {\bar z}}. \label{eqn:lndef}
\end{equation}
Note that despite of their benign appearance, the expressions in Eq.(\ref{eqn:lndef})  contain divergent poles for $n\leq-2$. However, including  these divergent terms is critical for the generators $l_{n} $ and $ {\bar l}_{n}$ to form the Witt algebras or the Virasoro algebras without the central extension:
\begin{equation}
[l_{n}, l_{m}] = (n-m)l_{n+m} \ , \  [{\bar l}_{n}, {\bar l}_{m}] = (n-m){\bar l}_{n+m}.
\end{equation}

\subsection{Generalization}\label{sec:generalization}
At this point, it would be beneficial to take notice of the fact that the choice of the specific transformations in Eq. (\ref{eqn:enn}) was, though it is in accordance with the Laurent series and certainly stands as a natural one, completely arbitrary. Therefore, we would like to reexamine the above procedure by introducing the following (holomorphic) differential operators, which is more general than those in Eq. (\ref{eqn:lndef}):
\begin{equation}
\l_{\kappa}=-g(z)f_{\kappa}(z)\frac{\partial}{\partial z}, \label{eqn:Lkdef}
\end{equation}
where $g(z)$ and $f_{\kappa}(z)$ are  both holomorphic functions on $z$, and $\kappa$ is the index that specifies the holomorphic function $f_{\kappa}(z)$. 
We  see in the following that for certain choices of $g(z)$, $f_{\kappa}(z)$ and the algebra formed by $\l_{\kappa}$ are consistently derived. In particular, choosing $g(z)=z$ reproduces the above argument that leads to the Witt algebra (the classical Virasoro).

First, we impose the following relation on   $f_{\kappa}(z)$:
\begin{equation}
\l_{0} f_{\kappa} (z)= -\kappa f_{\kappa}(z). \label{eqn:fkdef}
\end{equation}
As we shall shortly see, we can assume without inconsistency
\begin{equation}
\l_{0}=-g(z)\frac{\partial}{\partial z}, \label{eqn:L0def}
\end{equation}
or 
\begin{equation}
f_{0}(z)=1; \label{eqn:f0df}
\end{equation}
hence, Eq. (\ref{eqn:fkdef}) reads
\begin{equation}
g(z)\frac{\partial}{\partial z} f_{\kappa} (z)= \kappa f_{\kappa}(z). \label{eqn:fkderv}
\end{equation}

One can readily solve Eq. (\ref{eqn:fkderv}) as
\begin{equation}
f_{\kappa}(z)=A_{\kappa}e^{\kappa\int^{z}\frac{dz}{g(z)}}, \label{eqn:fksol}
\end{equation}
where $A_{\kappa}$ is a constant of integration. Note that Eq. (\ref{eqn:fksol}) yields $f_{0}(z)=1$ if we take $A_{0}$ to be unity, which we do for the rest of the paper.

The commutation relation between $\l_{\kappa}$ is 
\begin{equation}
[\l_{\kappa}, \l_{\kappa'}]=(\kappa' -\kappa)g(z)f_{\kappa}(z)f_{\kappa'}(z)\frac{\partial}{\partial z} \ ,
\end{equation}
where we utilized Eq. (\ref{eqn:fkderv}) in the following form:
\begin{equation}
\frac{\partial}{\partial z}f_{\kappa}(z)=\frac{\kappa}{g(z)}f_{\kappa}(z). \label{eqn:derfk}
\end{equation}
Noting that
\begin{equation}
f_{\kappa}(z)f_{\kappa'}(z)=A_{\kappa}A_{\kappa'}e^{(\kappa+\kappa')\int^{z}\frac{dz}{g}}=\frac{A_{\kappa}A_{\kappa'}}{A_{\kappa+\kappa'}}f_{\kappa+\kappa'}(z)
\end{equation}
from Eq. (\ref{eqn:fksol}), we arrive at the Witt algebra;
\begin{equation}
[\l_{\kappa}, \l_{\kappa'}]=
(\kappa -\kappa')
\l_{\kappa+\kappa'}, \label{eqn:Witt}
\end{equation}
if we impose $A_{\kappa}A_{\kappa'}=A_{\kappa+\kappa'}$, which is obviously satisfied by 
\begin{equation}
A_{k}=e^{{const.}\kappa}.
\end{equation}

Although we have introduced $f_{\kappa}(z)$ simply as a means to define $\l_{\kappa}$,  the action of $\l_{\kappa}$ on $f_{\kappa}(z)$ is of some interest.
To see this, we derive
\begin{equation}
\l_{\kappa}f_{\kappa'}(z)=f_{\kappa}(z)\l_{0}f_{\kappa'}(z)=-\kappa'f_{\kappa}(z)f_{\kappa'}(z)=-\kappa'f_{\kappa+\kappa'}(z).
\end{equation}
The action of $\l_{\kappa}$ on $f_{\kappa'}(z)$ alters the eigenvalue of $f_{\kappa'}(z)$ by the amount of $\kappa$ and multiplies by $-\kappa'$ to obtain $-\kappa' f_{\kappa+\kappa'}(z)$. Therefore,  the Witt algebra (\ref{eqn:Witt}) can be represented over the linear space spanned by $f_{\kappa}(z)$'s.

The analysis presented above can be repeated for the other set of differential operators, which are characterised by
\begin{equation}
{\bar \l}_{0}\equiv-g({\bar z})\frac{\partial}{\partial {\bar z}}, \label{eqn:L0bdf}
\end{equation}
where ${\bar z}$ stands for the complex conjugate of $z$. It is apparent then that $f_{\kappa}({\bar z})$ serves as a basis for the space over which the Witt algebra for $ {\bar \l}_{\kappa}$ is represented. Thus, we have constructed two independent sets of the Witt algebra and the representation space.

So far, we imposed no restriction on the nature of the index $\kappa$, which can take either discrete or continuous values, or even complex values. It turns out that the domain of $\kappa$ depends on our choice of $g(z)$. Because $\kappa$ is the index for the basis that spans the representation space,  the consideration of the representation imposes restrictions on $\kappa$. 

For the choice of $g(z)$, three cases are particularly interesting, each of which corresponds to one of three representative points of $SL(2, \mathbb{R})$ parameter space respectively, as explained in the introduction. The first one is
\begin{equation}
g(z)=z.
\end{equation}
This choice results in
\begin{equation}
\l_{0}=-z \frac{\partial}{\partial z}=l_{0}
\end{equation}
and
\begin{equation}
f_{\kappa}=\exp(\kappa\int^{z}\frac{dz}{z})e^{\kappa \cdot const.}=z^{\kappa}e^{\kappa \cdot const.}, \label{eqn:fkord}
\end{equation}
thus,
\begin{equation}
\l_{\kappa}=-z^{\kappa+1}\frac{\partial}{\partial z}
\end{equation}
up to the constant factor $e^{\kappa \cdot const.}$. Therefore, if we demand that the basis of the representation $f_{\kappa}$  be single-valued, $\kappa$ must take integer values; 
otherwise, $f_{\kappa}$ would produce cuts on the complex plane and  $f_{\kappa}(z)$ would be determined only up to some phase factor. \footnote{This feature can be exploited to accommodate fermions on the worldsheet.}
With this restriction on $\kappa$, the differential operators $\l_{n} \ (n \in \mathbb{Z})$ are nothing but the generators  $l_{n}$ in Eq. (\ref{eqn:lndef}). Here, we have simply established the usual set of the generators of the conformal transformation in two dimensions in the form of the differential operators and made the connection with the corresponding (classical) Virasoro algebra.

For the other choices, we encounter a novel situation.
Choosing 
\begin{equation}
g(z)=-\frac12 (z-1)^{2},  \label{eqn:ssdgz}
\end{equation}
yields
\begin{equation}
\l_{0}=-\left\{-\frac12 (z-1)^{2} \right\}\frac{\partial}{\partial z}=(-z+\frac{z^{2}+1}{2}) \frac{\partial}{\partial z}=l_{0}-\frac{l_{1}+l_{-1}}{2}\label{eqn:L0L1L-1}
\end{equation}
and
\begin{equation}
f_{\kappa}=\exp(\kappa\int^{z}\frac{2dz}{(z-1)^{2}})e^{\kappa \cdot const.}=\exp(-\frac{2\kappa}{z-1})e^{\kappa \cdot const.}. \label{eqn:fkssd}
\end{equation}
Equation (\ref{eqn:L0L1L-1}) corresponds to the `light-cone' or `massless representation', as observed in the introduction.
Note that, unlike Eq. (\ref{eqn:fkord}), $f_{\kappa}$ in Eq. (\ref{eqn:fkssd}) takes a single definite value everywhere save for at $z=1$, for any real $\kappa$. Thus, provided $\kappa$ is real, there is no restriction on $\kappa$, as expected from the reasoning presented in the introduction. The generators are now continuously indexed and take the form
\begin{equation}
\l_{\kappa}=-\frac12 (z-1)^{2}\exp(-\frac{2\kappa}{z-1})\frac{\partial}{\partial z}
\end{equation}
up to the $\kappa$-wise constant multiplication $e^{\kappa \cdot const.}$, which turns out to be irrelevant and is thus neglected by choosing the $const.$ to be nil.

For the rest of the presentation, we mostly treat these two generators and the corresponding conserved charges. However, in passing, the following third and last choice is worth mentioning: 
\begin{equation}
g(z)=z^{2}+1. \label{eqn:gzz2p1}
\end{equation}
This choice corresponds to 
\begin{equation}
\l_{0} \equiv l_{1}+l_{-1}, \label{eqn:K1def}
\end{equation}
and yields
\begin{equation}
f_{\kappa}=\exp(\kappa\int^{z}\frac{dz}{z^{2}+1})e^{\kappa \cdot const.}=\left(\frac{i+z}{i-z}\right)^{-\frac{i}{2}\kappa}e^{\kappa \cdot const.}=e^{\kappa \arctan z}e^{\kappa \cdot const.} 
.
\label{eqn:fk1-1}
\end{equation}
Thus,
\begin{equation}
\l_{\kappa}=-(z^{2}+1)e^{\kappa \arctan z}\frac{\partial}{\partial z}
\end{equation}
up to the $\kappa$-wise constant multiplication $e^{\kappa \cdot const.}$.

Note that $\arctan z$ is a function that yields multiple values that differ by 
$\pi n ( n \in \mathbb{Z})$. Demanding $f_{\kappa}$, which spans the representation space, to be single-valued would then restrict the value of $\kappa$ to be
\begin{equation}
\kappa = 2 \mathbb{Z} i\label{eqn:kiZ} 
\end{equation}
from the expression of $f_{\kappa}$.
Then, after proper redefinition, the Witt algebra which is defined by
$[\l'_{p}, \l'_{q}]=(p-q)\l'_{p+q}  (p,q \in \mathbb{Z})$ emerges for this case.
The result here, particularly the appearance of the imaginary number in Eq. (\ref{eqn:kiZ}), seems consistent with the intuition presented in the introduction because Eq. (\ref{eqn:K1def}) can be related to  a `space-like'  representation. 
%
%
%
%
%
%
%
%

\subsection{Dipolar quantization}

In the analysis that established the generators $\l_{\kappa}$, the $\l_{0}$ generator, among others, was treated singularly. This is justified by the fact that  $L_{0}$ generator, which is $\l_{0}$ for the choice of $g(z)=z$,  actually corresponds to  the time and space translation and plays a singular role. In this subsection, we would like to generalise this feature to general $\l_{0}$. We should then be able to adopt, for example, $L_{0}-L_{+}$ as  (the holomorphic part of) the Hamiltonian.

Following the case for $L_{0}$ (and ${\bar L}_{0}$), let us define the time coordinate $t$ by
\begin{equation}
-\frac{\partial}{\partial t} \equiv \l_{0} +{\bar  \l}_{0}
\end{equation}
and the space coordinate $s$ by
\begin{equation}
-\frac{\partial}{\partial s} \equiv i \left( \l_{0} -{\bar  \l}_{0}\right).
\end{equation}
By using Eqs. (\ref{eqn:L0def}) and (\ref{eqn:L0bdf}), these two equations can be summarised in the following matrix form:
\begin{equation}
\left(\begin{array}{c}\frac{\partial}{\partial t} \\\frac{\partial}{\partial s}\end{array}\right) =\left(\begin{array}{cc}g(z) & g({\bar z}) \\ig(z) & -ig({\bar z})\end{array}\right)\left(\begin{array}{c}\frac{\partial}{\partial z} \\\frac{\partial}{\partial {\bar z}}\end{array}\right). \label{eqn:jacobitszzbar}
\end{equation}
The inverse of the matrix in Eq. (\ref{eqn:jacobitszzbar}) can be read as the Jacobian matrix for $t(z, {\bar z})$ and $s(z, {\bar z})$ 
\begin{equation}
\left(\begin{array}{cc}g(z) & g({\bar z}) \\ig(z) & -ig({\bar z})\end{array}\right)^{-1}=\left(\begin{array}{cc}\frac{\partial t}{\partial z}& \frac{\partial s}{\partial z} \\\frac{\partial t}{\partial {\bar z}} & \frac{\partial s}{\partial {\bar z}}\end{array}\right).
\end{equation}
Thus, the following sets of equations are obtained:
\begin{equation}
\left\{\begin{array}{c} \frac{\partial t}{\partial z}=\frac{1}{2g(z)} \\   \frac{\partial s}{\partial z}=\frac{1}{2ig(z)}        \end{array}\right.
\end{equation}
and
\begin{equation}
\left\{\begin{array}{c}\frac{\partial t}{\partial {\bar z}}=\frac{1}{2g(\bar {z})}  \\ \frac{\partial s}{\partial {\bar z}}=-\frac{1}{2ig(\bar {z})} \end{array}\right. .
\end{equation}
Each equation can be readily integrated as
\begin{equation}
\left\{\begin{array}{c} t=\frac12 \int^{z}\frac{dz}{g(z)} +{\bar v}({\bar z})\\  is=\frac12 \int^{z}\frac{dz}{g(z)} +{\bar v'}({\bar z}) \end{array}\right. \label{eqn:t/g}
\end{equation}
and
\begin{equation}
\left\{\begin{array}{c}t=\frac12 \int^{{\bar z}}\frac{d{\bar z}}{g({\bar z})} +v'({z})\\ is=-\frac12 \int^{{\bar z}}\frac{d{\bar z}}{g({\bar z})} +v({z})\end{array}\right. , \label{eqn:s/g}
\end{equation}
where $v$, $v'$ ($\bar v$, $\bar v'$) can be any holomorphic (anti-holomorphic) functions.
Combining Eqs. (\ref{eqn:t/g}) and (\ref{eqn:s/g}), it is apparent that
\begin{eqnarray}
t=\frac12 \left( \int^{z}\frac{dz}{g(z)} +\int^{{\bar z}}\frac{d{\bar z}}{g({\bar z})}\right) , \\
is=\frac12\left( \int^{z}\frac{dz}{g(z)} -\int^{{\bar z}}\frac{d{\bar z}}{g({\bar z})}\right) .
\end{eqnarray}
Note that the following combination yields a particularly simple expression as
\begin{equation}
t+is=\int^z \frac{dz}{g(z)}; \label{eqn:tisig}
\end{equation}
hence, we  denote the combination $t+is$ as a single complex variable $w$, which is used in the latter part of the paper.

Let us be  specific  with the case $g(z)=z$. Then
\begin{equation}
\int^{z}\frac{dz}{g(z)}= \ln z, \label{eqn:intzgz}
\end{equation}
so
\begin{eqnarray}
t=\frac12 \left( \ln z + \ln {\bar z}\right) =\frac12 \ln (z{\bar z}), \\
is=\frac12\left( \ln z - \ln {\bar z}\right) =\frac12 \ln \frac{z}{\bar z} .
\end{eqnarray}
These two equations can be concisely summarised as
\begin{equation}
t+is=\ln z.
\end{equation}
Thus, in the radial coordinate $z=re^{i\phi}$,
\begin{equation}
t=\ln r, \quad
is=i\phi , \label{eqn:tsrphi}
\end{equation}
as is well known. This relationship between  time translation and radial translation is the origin of the term `radial quantization' \cite{Fubini:1972mf}.

Let us now examine the case of interest:  $g(z)=-\frac12(z-1)^{2}$. Starting from
\begin{equation}
\int^{z}\frac{dz}{g(z)}= \frac{2}{z-1}
\end{equation}
we obtain
\begin{eqnarray}
t=\frac12 \left( \frac{2}{z-1} + \frac{2}{{\bar z}-1}\right) , \\
is=\frac12\left(  \frac{2}{z-1} - \frac{2}{{\bar z}-1}\right) ;
\end{eqnarray}
hence, it would be useful to consider the combination
\begin{equation}
t+is=\frac{2}{z-1}. \label{eqn:tszdi}
\end{equation}
Reciprocally, 
\begin{equation}
z=1+\frac{2}{t+is}
\end{equation}
and, if we introduce the Cartesian coordinate for $z=x+iy$, we obtain
\begin{equation}
z=x+iy=1+\frac{2}{t+is}=1+\frac{2(t-is)}{t^{2}+s^{2}},
\end{equation}
thus arriving at
\begin{equation}
\left\{\begin{array}{c} x=1+\frac{2t}{t^{2}+s^{2}}\\ y=-\frac{2s}{t^{2}+s^{2}}\end{array}\right. . \label{eqn:xyts}
\end{equation}

\begin{figure}[tbh]
\centering
\includegraphics[width=12.5cm]{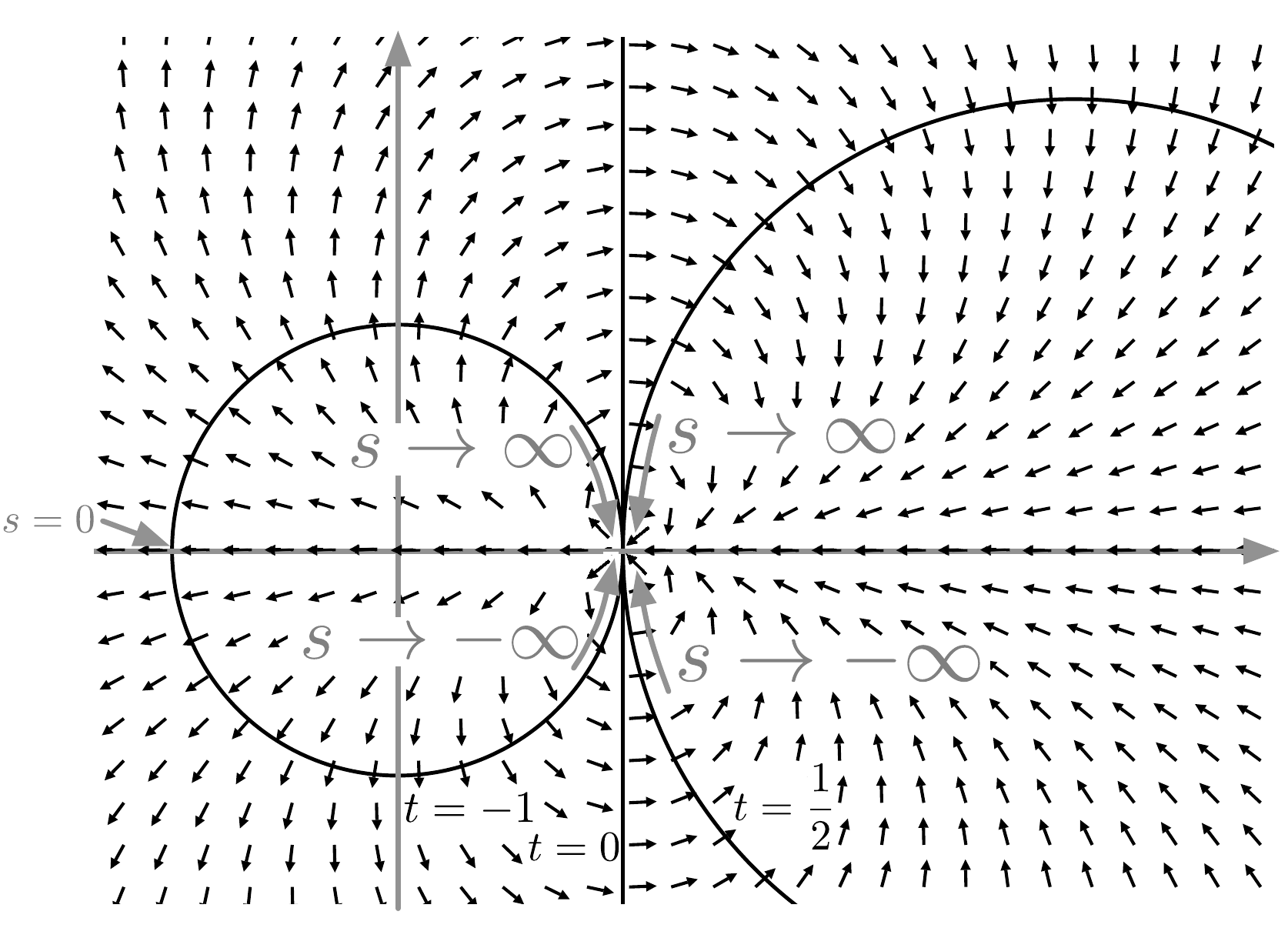}
\caption{Vector field of time translation generated by Eq. (\ref{eqn:dpltmtr}) is depicted along with the equal-time contours for $t=-1, 0$, $\frac12$. All the equal-time contours contact each other at $z=1$, where a fictitious dipole is located.}
\label{fig:dipolar}
\end{figure}

From Eq. (\ref{eqn:xyts}),  the following combination produces the simple expression
\begin{equation}
\left(x-1-\frac{1}{t} \right)^{2}+y^{2}=\frac{(t^{2}-s^{2})^{2}}{t^{2}(t^{2}+s^{2})^{2}}+\frac{4s^{2})^{2}}{t^{2}(t^{2}+s^{2})^{2}}=\frac{1}{t^{2}},
\end{equation}
which is nothing but a circle of radius $\frac{1}{|t|}$. Therefore, the contour of the constant $t$ depicts a circle of radius $\frac{1}{|t|}$ with the center located at $\left( 1+\frac{1}{t}\right)$. While these circles  have different radii for different values of $t$, they always go through the point $z=1$, as illustrated in Fig. \ref{fig:dipolar}. In particular, the contours for $t=-\infty$ and $t=\infty$ are degenerate at the point $z=1$. If we increase $t$ starting from $t=-\infty$, we have circles on the left of $z=1$ until we reach $t=0$. At that point, the contour becomes the great circle that  evenly intersects the entire Riemann sphere. This great circle renders a straight line in the complex plane.
For $t>0$, the contour circles are on the right of $z=1$ and converge to $z=1$ as $t\rightarrow\infty$.

Along each contour circle of the constant $t$, the parameter $s$ increases from $s=-\infty$ at $z=1$ as it covers the lower half of the circle and reaches $s=0$ at the crossing with the real axis. Increasing $s$ further sweeps the upper half of the circle and eventually encircles back to $z=1$ as $s\rightarrow\infty$. This is contrasted with the ordinary CFT analysis where the spatial parameter $s$ has a finite range, as is apparent from Eq. (\ref{eqn:tsrphi}). The emergence of the infinite range in the space parametrization is consistent with the continuous index for the classical Virasoro algebra and also suggests the continuous spectrum.

If we formulate the time-translational vector field in  Cartesian coordinates, we find
\begin{equation}
\frac{\partial}{\partial t}=\frac{-(x-1)^{2}+y^{2}}{2}\frac{\partial}{\partial x}-(x-1)y\frac{\partial}{\partial y}. \label{eqn:dpltmtr}
\end{equation}
The above vector field produces the dipole field whose origin is located at $z=1$ and points along the real axis (Fig. \ref{fig:dipolar}) .
In contrast with radial quantization \cite{Fubini:1972mf}, we regard this translation generated by  Eq. (\ref{eqn:dpltmtr}) as the time translation associated with the `dipolar' quantization \cite{Ishibashi:2015jba}, after the configuration of the time-translational vector field.
\footnote{Other than radial quantization, there is another quantization called N-S (North-South pole) quantization, found in the literature \cite{Luscher:1974ez,Rychkov:2016iqz}. 
}

The differential operator (\ref{eqn:dpltmtr}) can also be written in the form
\begin{equation}
{\hat D}-\frac{{\hat K}_{1}+{\hat P}_{1}}{2},  \label{eqn:dplDKP}
\end{equation}
by using the generators ${\hat D}$,  ${\hat P}_{\mu}$ and ${\hat K}_{\mu}$, which correspond to dilation,  translation and  special conformal translation, respectively:
\begin{eqnarray*}
 \hbox{[dilation]} \quad \qquad  {\hat D}&=&x^{\mu}\partial_{\mu} \ ,\\
  \hbox{[translation]} \ \  \quad  {\hat P}_{\mu}&=&\partial_{\mu}\ ,\\
 \hbox{[SCT]}\qquad\qquad{\hat K}_{\mu} &=&2x_{\mu}x^{\nu}\partial_{\nu}-(x\cdot x)\partial_{\mu}\ .
\end{eqnarray*}
Equation (\ref{eqn:dplDKP}) is  especially useful when we  extend the current result to other dimensions. In addition, it is easy to see that another direction exits where we can consider a similar differential operator ${\hat D}-({\hat K}_{2}+{\hat P}_{2})/{2}$. This would generate a vector field that is simply the ninety degree rotation of the vector field (\ref{eqn:dpltmtr}).

To conclude our geometrical analysis, we comment on the significance of the result.
To study  the corresponding quantum system based on the aforementioned geometrical analysis, we need to examine  the conserved charges generated by the conformal transformation
\begin{equation}
Q^{\epsilon}=\int dx J_{0}^{\epsilon}. \label{eqn:QeJ}
\end{equation}
The integral in the equation above should be performed along a loop on which `time' is constant. Conservation of the current $J^{\epsilon}$ ensures that the integral does not depend on the choice of the loop or the value of the time on the loop. Thus,  identifying  the time translation is crucial.

\section{Quantum analysis}\label{sec:QA}
\subsection{The Virasoro algebra}\label{subsec:Va}
In the previous section, we  systematically analyzed the time translation evoked by a class of Hamiltonians that includes the ordinary [$L_{0}$] and the SSD [$L_{0}-(L_{1}+L_{-1})/2$] Hamiltonians.
Now that we understand the nature of these time translations, let us proceed to study the conserved charges associated with each of them.
Consider a conserved charge, which is the integration of a current that is associated with a certain coordinate transformation $\epsilon$ [Eq. (\ref{eqn:zezbeb})], as shown in Eq. (\ref{eqn:QeJ}).  Observing the relationship between the current and the energy-momentum tensor [Eq. (\ref{eqn:JTe})], we arrive at the following:
\begin{equation}
Q^{\epsilon}=\int dx J_{0}^{\epsilon} =\frac{1}{2\pi i}\oint \kern-0.7em \ ^{t}\kern.2em dz \epsilon(z) T_{zz}(z), \label{eqn:QeJeT}
\end{equation}
where $t$ is attached to the sign of the integral as a reminder of the fact that the integral should be performed along a  path with the constant $t$, and a natural normalization factor $2\pi i$ is included.
In the analysis of the previous section,  $\epsilon(z)$ was further divided into the product of $g(z)$ and $f_{\kappa}$ [see Eq. (\ref{eqn:Lkdef})]. In this case, the conserved charges labelled by $\kappa$ are determined in the following way:
\begin{equation}
\L_{\kappa} \equiv \frac{1}{2\pi i}\oint \kern-.7em \ ^{t}\kern.4em  dz g(z)f_{\kappa}(z)T(z), \label{eqn:cLkapdf}
\end{equation}
where $T(z)=T_{zz}(z)$ is the holomorphic part of the energy-momentum tensor of the original CFT. 
In particular, for $\kappa =0$, we have the expression for the Hamiltonian for each set of $g(z)$ and $f_{\kappa}(z)$ corresponding to either the case for  radial quantization or for  dipolar quantization:
\begin{equation}
\hspace{-2em} 
\L_{0}=\frac{1}{2\pi i}\oint
\kern-.5em \ ^{t}\kern.4em  dz g(z)T(z)=\left\{\begin{array}{ll}L_{0} &  \hbox{for} \ g(z)=z \hbox{ [radial]} \\ \ & \ \\L_{0}-\frac{L_{1}+L_{-1}}{2} & \hbox{for} \ g(z)=-\frac{(z-1)^2}{2}\hbox{ [dipolar]} \end{array}\right.,   \label{eqn:cL0df}
\end{equation}
keeping Eq. (\ref{eqn:f0df}) in mind.

\begin{figure}[tbh]
\centering
\includegraphics[width=14cm]{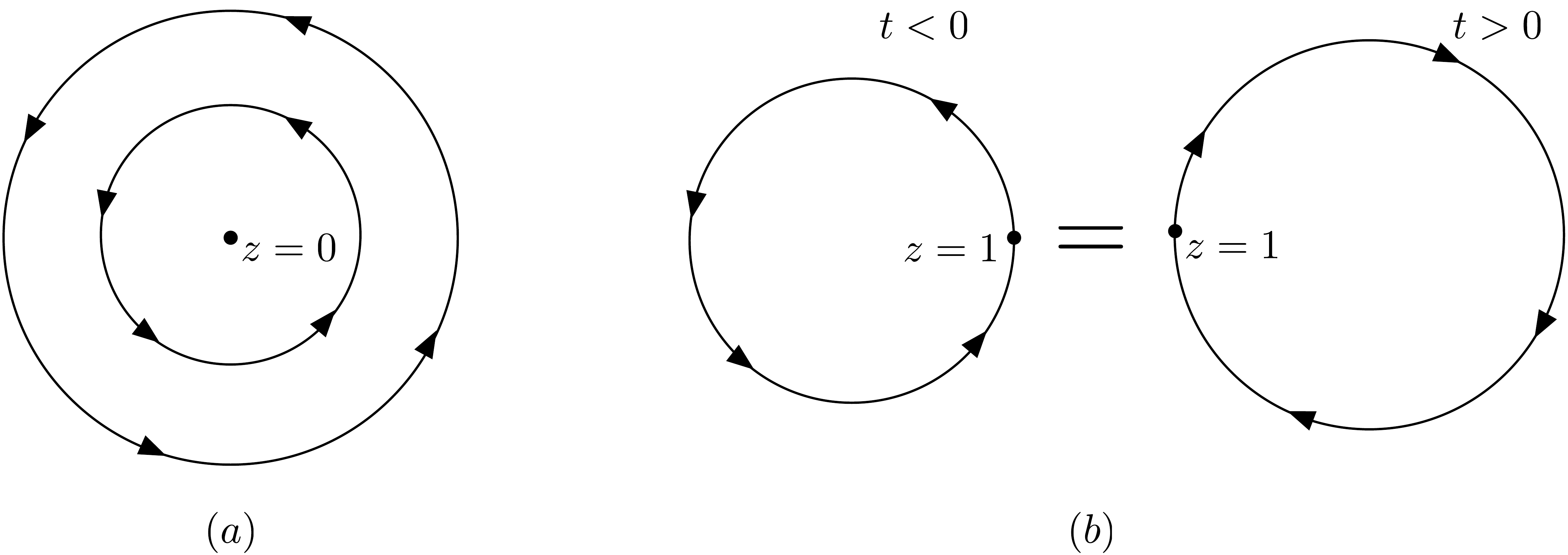}
\caption{Equal-time paths are depicted for $ (a) $ radial quantization and $ (b) $ dipolar quantization. The apparent direction of the contour depends on the sign of $t$ for  dipolar quantization.}
\label{fig:equaltimepath1}
\end{figure}
We now clarify a point that has not been made specific. When the integral along the path of constant time was introduced in Eq. (\ref{eqn:QeJeT}), we did not specify the direction of the integral along the path, which can be in one of two directions.  The first direction is  for ordinary radial quantization, in which case the contour integral is done in the  counter-clockwise direction [Fig. \ref{fig:equaltimepath1} $(a)$]. In terms of the parameters $t$ and $s$, this integral is done from $s=0$ to $s=2\pi$:
\begin{equation}
Q=\frac{1}{2\pi i}\oint^{s=2\pi}_{s=0} \kern-2.4em \ ^{t}\kern1.em dz q(z).
\end{equation}

\begin{figure}[tbh]
\centering
\includegraphics[width=10cm]{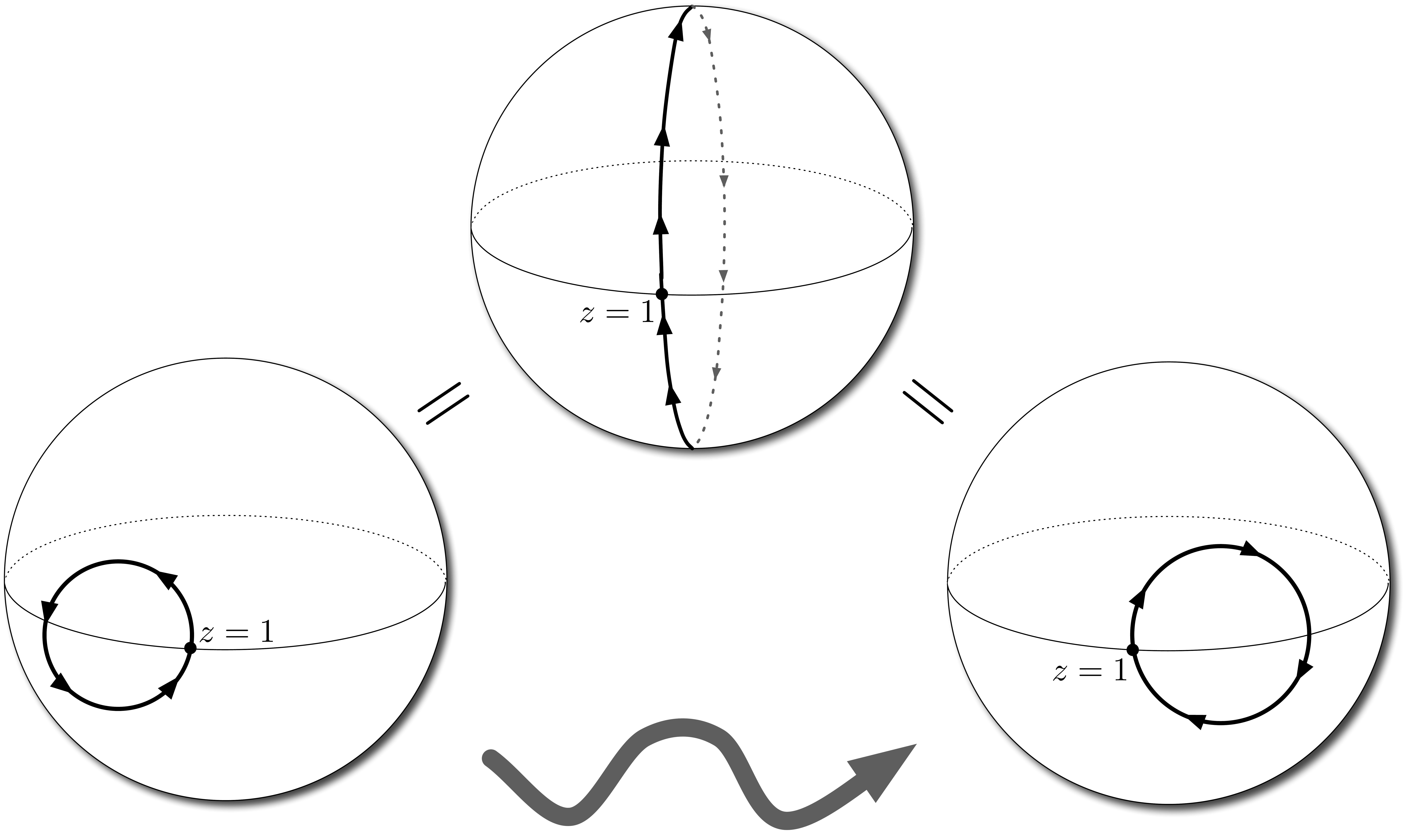}
\caption{Contour integral for $t <0$ is converted to the integral for $t>0$, switching the apparent directions of the integration.}
\label{fig:contouronsphere}
\end{figure}

\begin{figure}[tbh]
\centering
\includegraphics[width=16.5cm]{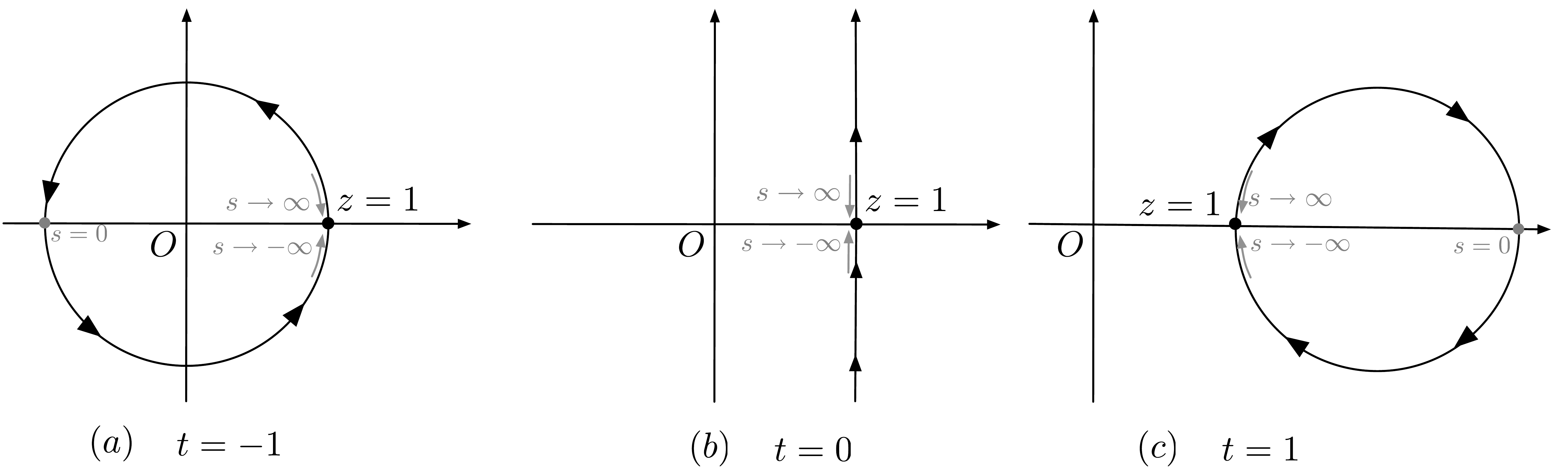}
\caption{Equal-time paths for $(a)$ $t=-1$, $(b)$ $t=0$, and $(c)$ $t=1$.}
\label{fig:equaltimepath2}
\end{figure}

The second case requires more caution. For this case, we integrate  in the  counter-clockwise direction, if $t<0$, but in the clockwise for $t>0$ [Fig. \ref{fig:equaltimepath1} $(b)$]. Although these two contours in Fig. \ref{fig:equaltimepath1} $(b)$ imply the integrations in opposite directions, they yield the same value. This fact can be understood by deforming the contour continuously from $t<0$ to $t>0$. During this deformation, the contour passes through the infinity point, which is located at the antipode of the origin on the Riemann sphere (Fig. \ref{fig:contouronsphere}).
In terms of the $t, s$ parameters, the integration is done from $s=-\infty$ to $s=\infty$, though it may look awkward:
\begin{equation}
Q=\frac{1}{2\pi i}\oint^{s=-\infty}_{s=\infty} \kern-2.8em \ ^{t}\kern1.4em dz q(z).
\end{equation}
Figure \ref{fig:equaltimepath2} shows the equal-time paths on the complex plane for $t=-1, 0, 1$. Setting $t=-1$ gives the integral over the unit circle just as for the case of $t=0$ in  radial quantization (\ref{eqn:tsrphi}). The case for $t=0$, as in Fig. \ref{fig:equaltimepath2} $(b)$, is particularly interesting because the path apparently becomes a straight line, which facilitates calculations as we will see shortly.

\begin{figure}[tbh]
\centering
\includegraphics[width=15cm]{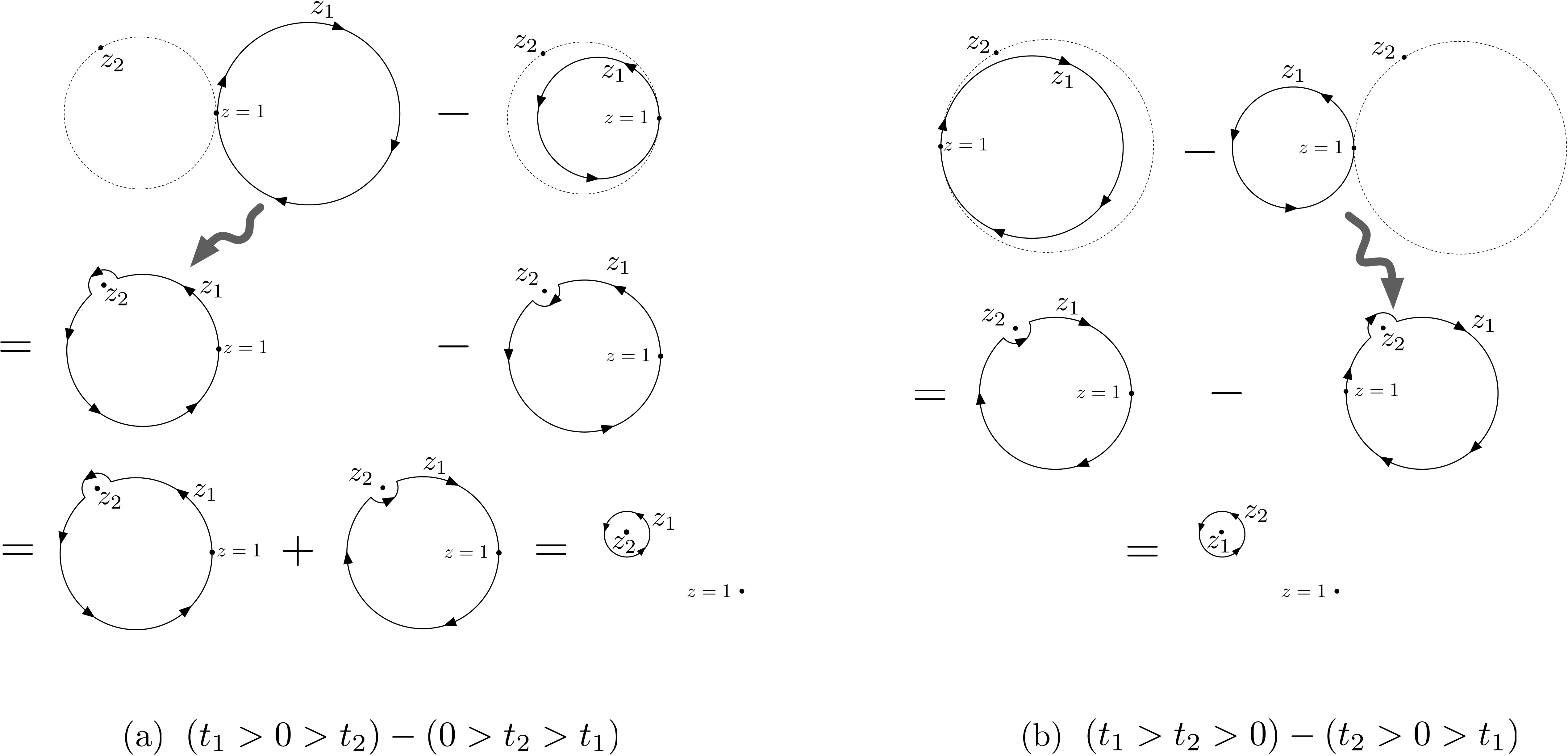}
\caption{The subtractions between two integrals explained pictorially. Different choices of $t_{1}$ and $t_{2}$ amount to the same contour integral around $z_{2}$. }
\label{fig:contoursubtraction}
\end{figure}

We now calculate the commutation relations between the following more general quantities:
\begin{equation}
Q^{(i)}=\frac{1}{2\pi}\oint \kern-0.7em \ ^{t_{i}}\kern0.2em dz_{i} q^{(i)}(z_{i}).
\end{equation}
Assuming $q^{(i)}(z_{i})$ is an operator in quantum field theory, we introduce the symbol for the time-ordered product $\mathbf{T}$ as
\begin{equation}
\mathbf{T}\left(q^{(1)}(z_{1}) q^{(2)}(z_{2}) \right)=\left\{\begin{array}{cc} q^{(1)}(z_{1}) q^{(2)}(z_{2})& \hbox{for } t_{1}>t_{2} \\q^{(2)}(z_{2})q^{(1)}(z_{1})  &  \hbox{for }  t_{1}<t_{2}\end{array}\right. .
\end{equation}
Then,
\begin{eqnarray}
\left[ Q^{(1)}, Q^{(2)} \right]&&=\left[ \frac{1}{2\pi i} \oint \kern-.6em \ ^{t_{1}}\kern.2em dz_{1}q^{(1)}(z_{1}),  \frac{1}{2\pi i}   \oint \kern-.6em \ ^{t_{2}}\kern.2em dz_{2}q^{(2)}(z_{2}) \right]  \nonumber \\
&&= \frac{1}{2\pi i}   \oint \kern-.6em \ ^{t_{2}}\kern.2em dz_{2} {\Biggl ( }
\frac{1}{2\pi i}  \int_{\includegraphics[width=3em]{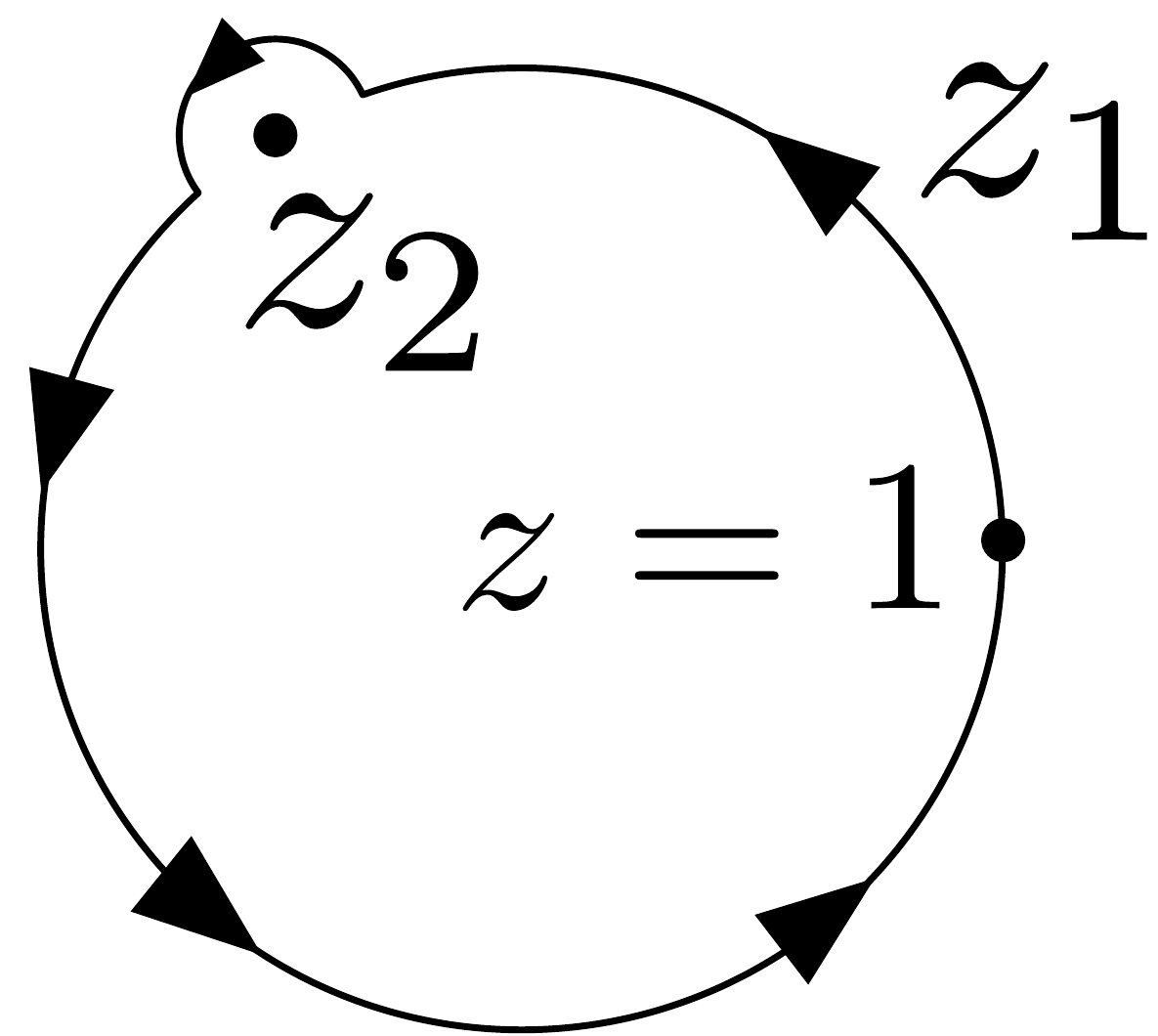}} dz_{1} q^{(1)}(z_{1}) q^{(2)}(z_{2}) \nonumber \\
&&\qquad\qquad\quad-\frac{1}{2\pi i}  \int_{\includegraphics[width=3em]{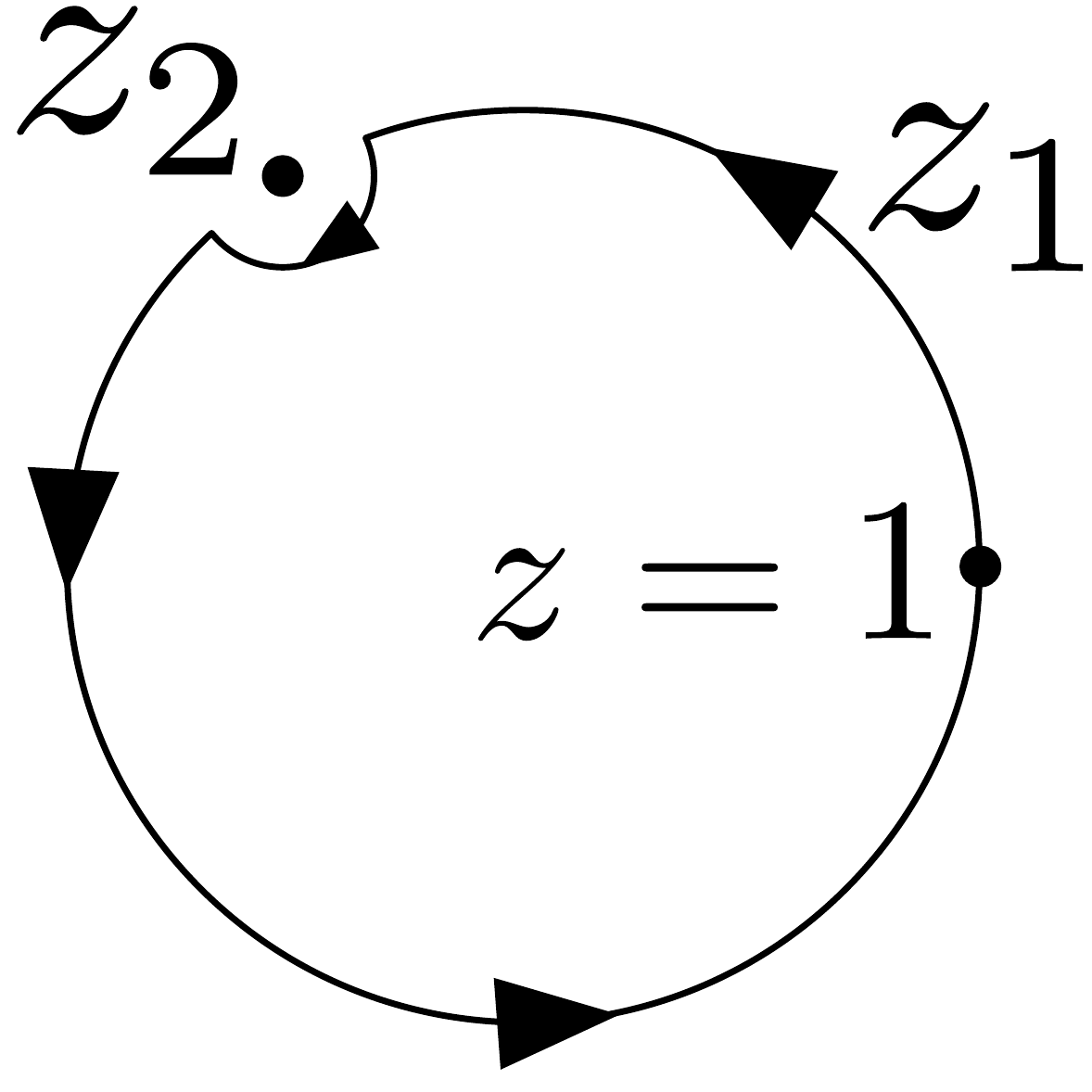}} dz_{1}  q^{(2)}(z_{2})q^{(1)}(z_{1})
\Biggr) \label{eqn:QQcom} \\
&&= \frac{1}{2\pi i}   \oint \kern-.6em \ ^{t_{2}}\kern.2em dz_{2} {\Biggl [ }
 \Big( \frac{1}{2\pi i}  \int_{\includegraphics[width=3em]{z1z2contour.pdf}} dz_{1} - \int_{\includegraphics[width=3em]{z2z1contour.pdf}} dz_{1}\Big) \nonumber
 \\&& \qquad \qquad \qquad  \qquad  \qquad \times \mathbf{T}\left(q^{(1)}(z_{1}) q^{(2)}(z_{2}) \right) \Biggl ] \nonumber \\
 &&=\frac{1}{2\pi i}   \oint \kern-.6em \ ^{t_{2}}\kern.2em dz_{2} \int_{\includegraphics[width=1.5em]{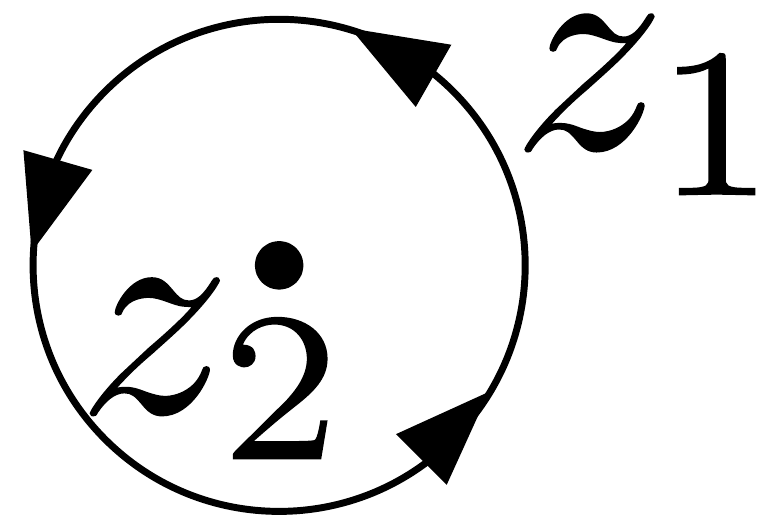}} dz_{1} \mathbf{T}\left(q^{(1)}(z_{1}) q^{(2)}(z_{2})  \right) . \nonumber
\end{eqnarray}
Here, we understand that $t_{1}$ is set to be $-1$ for  dipolar quantization and to be $0$ for radial quantization so that the contour  for $z_{1}$ becomes the unit circle. The value $t_{2}$ is also chosen appropriately in each term. The contour for $z_{1}$ is  shown pictorially in  Eq.(\ref{eqn:QQcom}). One could choose different values for $t_{1}$ and $t_{2}$  and yet obtain the same result (Fig. \ref{fig:contoursubtraction}).

The commutation relations between the above-defined conserved charges can be derived straightforwardly by using the following operator product expansion of the energy-momentum tensor:
\begin{equation}
 \mathbf{T}\left( T(z_{1})T(z_{2}) \right)\sim \frac{c/2}{(z_{1}-z_{2})^{4}}+\frac{2T(z_{2})}{(z_{1}-z_{2})^{2}} + \frac{\partial_{z}T(z)|_{z=z_{2}}}{z_{1}-z_{2}} + \cdots ,
\end{equation}
where $c$ is the central charge of the CFT.
From Eqs. (\ref{eqn:cLkapdf}) and (\ref{eqn:QQcom}), we have
\begin{eqnarray}
[\L_{\kappa} , \L_{\kappa'} ]&=&\frac{1}{2\pi i}   \oint \kern-.6em \ ^{t_{2}}\kern.2em dz_{2} g(z_{2})f_{\kappa'}(z_{2})\int_{\includegraphics[width=1.5em]{z1aroundz2.pdf}} dz_{1}g(z_{1})f_{\kappa}(z_{1}) \mathbf{T}\left(T(z_{1}) T(z_{2})  \right) \nonumber \\
&=& \frac{1}{2\pi i}   \oint \kern-.6em \ ^{t_{2}}\kern.2em dz_{2} g(z_{2})f_{\kappa'}(z_{2})\int_{\includegraphics[width=1.5em]{z1aroundz2.pdf}} dz_{1}g(z_{1})f_{\kappa}(z_{1})  \label{eqn:LLcom}\\
&& \times \left(  \frac{c/2}{(z_{1}-z_{2})^{4}}+\frac{2T(z_{2})}{(z_{1}-z_{2})^{2}} + \frac{\partial_{z}T(z)|_{z=z_{2}}}{z_{1}-z_{2}} \right).\nonumber 
\end{eqnarray}
After using the residue theorem and partial integrals, we arrive at
\begin{eqnarray}
[\L_{\kappa} , \L_{\kappa'} ]&=& \frac{c}{12}    \oint \kern-.6em \ ^{t_{2}}\kern.2em \frac{dz_{2} }{2\pi i}
\left\{ g \frac{\partial^{3}g}{\partial z^{3}}+
\kappa\left( 2\frac{\partial^{2}g}{\partial z^{2}}-\frac{1}{g}\left(\frac{\partial g}{\partial z}\right)^{2}\right)+\frac{\kappa^{3}}{g}\right\}f_{\kappa+\kappa'}(z_{2}) \nonumber \\
&&+ (\kappa-\kappa')    \oint \kern-.6em \ ^{t_{2}}\kern.2em \frac{dz_{2} }{2\pi i} g(z_{2})f_{\kappa+\kappa'}(z_{2})T(z_{2}) \\
&=& \frac{c}{12}    \oint \kern-.6em \ ^{t_{2}}\kern.2em \frac{dz_{2} }{2\pi i}
\left\{ g \frac{\partial^{3}g}{\partial z^{3}}+
\kappa\left( 2\frac{\partial^{2}g}{\partial z^{2}}-\frac{1}{g}\left(\frac{\partial g}{\partial z}\right)^{2}\right)+\frac{\kappa^{3}}{g}\right\}f_{\kappa+\kappa'}(z_{2}) \nonumber \\
&&+ (\kappa-\kappa')  \L_{\kappa+\kappa'}.\nonumber
\end{eqnarray}
This derivation also took advantage of Eq. (\ref{eqn:derfk}).

Let us now evaluate the term proportional to the central charge $c$. For further evaluation, we need the explicit form of $g(z)$. The explicit expression for $g(z)$ readily yields
\begin{eqnarray}
\hspace{-2em} 
&&\frac{c}{12}    \oint \kern-.6em \ ^{t_{2}}\kern.2em \frac{dz_{2} }{2\pi i}
\left\{ g \frac{\partial^{3}g}{\partial z^{3}}+
\kappa\left( 2\frac{\partial^{2}g}{\partial z^{2}}-\frac{1}{g}\left(\frac{\partial g}{\partial z}\right)^{2}\right)+\frac{\kappa^{3}}{g}\right\}f_{\kappa+\kappa'}(z_{2})\nonumber \\
&&=\left\{\begin{array}{ll}\frac{c}{12}    \oint \kern-.6em \ ^{\scriptscriptstyle t_{2}}\kern.2em \frac{dz_{{2}} }{2\pi i} 
\left\{ \frac{\kappa^{3}}{g}-\frac{\kappa}{g}\right\}f_{\kappa+\kappa'}(z_{2})
 &  \hbox{for} \ g(z)=z \hbox{ [radial]} \\ \ & \ \\
 \frac{c}{12}    \oint \kern-.6em \ ^{\scriptscriptstyle t_{2}}\kern.2em \frac{dz_{2} }{2\pi i}
\frac{\kappa^{3}}{g}f_{\kappa+\kappa'}(z_{2})
  & \hbox{for} \ g(z)=-\frac{(z-1)^2}{2}\hbox{ [dipolar]} \end{array}\right..  \label{eqn:ctmrd}
\end{eqnarray}
Besides the difference shown above, and that is caused by the explicit form of $g(z)$, note that $f_{\kappa}$ or the nature of $\kappa$ need not be specified until this point.

We now introduce a formula that shall prove useful for the further evaluation of Eq.(\ref{eqn:ctmrd}):
\begin{equation}
\oint \kern-.6em \ ^{t}\kern.2em   \frac{dz}{2\pi i}\frac{f_{\kappa}(z)}{g(z)}=\left\{\begin{array}{ll} \delta_{\kappa, 0}&  \hbox{for} \ g(z)=z \\ \ & \ \\ \delta(\kappa) & \hbox{for} \ g(z)=-\frac12(z-1)^2\end{array}\right.,    \label{eqn:Deltaf}
\end{equation}
where $\delta_{\kappa,0}$ is the Kronecker delta and $\delta(\kappa)$ is the Dirac delta function. The proof of this formula for $g(z)=z$   amounts to a simple residue integral. However, the case $g(z)=-\frac12(z-1)^2$ deserves an additional expounder. To facilitate the calculation, we set $t=0$ [see Fig. \ref{fig:equaltimepath2} $(b)$ for the integration path]. The equal-time path can then be parametrised as $z=1+iu, -\infty < u < \infty$. Noting that $f_{\kappa}=\exp (\frac{2}{z-1}\kappa)$ for this case, we obtain
\begin{equation}
\oint \kern-.6em \ ^{t}\kern.2em \frac{dz}{2\pi i}\frac{f_{\kappa}(z)}{g(z)}=\int^{\infty}_{-\infty}\frac{idu}{2\pi i} \frac{\exp(-i\frac{2\kappa}{u})}{u^{2}}.
\end{equation}
Changing variables to $x=-\frac{2}{u}$, the above integral yields
\begin{equation}
\frac{1}{2\pi}\int^{\infty}_{-\infty}dxe^{i\kappa x}
\end{equation}
which is nothing but the Dirac delta function; {\it quod erat demonstrandum}.

Combining all these calculations together, the result may be summarised as
\begin{equation}
[\L_{\kappa} , \L_{\kappa'} ]=(\kappa - \kappa')\L_{\kappa+\kappa'}
+\left\{\begin{array}{ll} \frac{c}{12}(\kappa^{3}-\kappa)\delta_{\kappa+\kappa', 0}&  \hbox{for} \ g(z)=z \\ \ & \ \\ \frac{c}{12}\kappa^{3}\delta(\kappa+\kappa') & \hbox{for} \ g(z)=-\frac12(z-1)^2\end{array}\right.. \label{eqn:Virkap}
\end{equation}
For $g(z)=z$, $\kappa$ is restricted to an integer if we demand that $f_{\kappa}$ be single valued in (\ref{eqn:fkord}); thus, the algebra generated by $\L_{\kappa}$'s becomes isomorphic to the ordinary Virasoro algebra generated by $L_{n}$'s. For dipolar quantization, where $g(z)=-\frac12(z-1)^2$, there is no such restriction applies to $\kappa$; it can be any real number. 

Note that the $\kappa$ coefficients  that immediately follow $c/12$ appear to differ between the radial and dipolar treatment in Eq. (\ref{eqn:Virkap}); Two comments on this fact are in order. First,  the term $\frac{c}{12}\kappa^{3}\delta(\kappa+\kappa')$  can be transformed to $\frac{c}{12}(\kappa^{3}-\kappa)\delta(\kappa+\kappa')$ by shifting $\L_{\kappa}$ as follows:
\begin{equation}
\L_{\kappa} \rightarrow \L_{\kappa} +\frac{c}{24}\delta(\kappa).
\end{equation}
The additional shift factor $\frac{c}{24}\delta(\kappa)$ can be also derived in the following way: Consider the 
 Schwarzian derivative term
\begin{equation}
\frac{c}{12}\frac{1}{(\partial_{z}u)^{2}}\left\{\left(\partial_{z}u\right)\left(\partial_{z}^{3}u\right)-\frac32\left(\partial_{z}^{2}u\right)^{2}\right\} =-\frac{2}{(z-1)^{4}}\cdot\frac{c}{12} \label{eqn:c24}
\end{equation}
which is associated with the non-trivial coordinate transformation
\footnote{The significance of this transformation is recognised in Ref. \cite{Tada:2014kza}. See also  \ref{sec:HTDO}.}
\begin{equation}
z\rightarrow u=e^{\frac{2}{z-1}}. \label{eqn:wez}
\end{equation}
Then integrate the result (\ref{eqn:c24}) along with $g(z)$ and $f_{\kappa}$, obtaining
\begin{equation}
\frac{c}{12}\oint \kern-.6em \ ^{t}\kern.2em \frac{dz}{2\pi i}{g(z)}{f_{\kappa}(z)}\frac{2}{(z-1)^{4}}=\frac{c}{24}\delta(\kappa).
\end{equation}
Therefore, the apparent dissimilarity between the $\kappa$ coefficients can be understood by the nontrivial difference between the coordinate systems. Second, the expression $\frac{c}{12}(\kappa^{3}-\kappa)\delta_{\kappa+\kappa', 0}$ suggests that the choice of values $\kappa=-1, 0, 1$ plays a special role. In fact, $L_{-1}, L_{0}$ and $L_{1}$ all annihilate the vacuum $|0\rangle$ in the case of the ordinary Virasoro algebra, as is well known. However, save for $\kappa=0$, there is no significant peculiarity at the values $\kappa=-1,1$ in the continuous Virasoro algebra; thus, $\frac{c}{12}\kappa^{3}\delta(\kappa+\kappa')$ appears to be more natural expression for the continuous Virasoro algebra.

%
%
One can also express $\L_{\kappa}$ in terms of the coordinate $w (=t+is)$. First, by combining Eqs. (\ref{eqn:fksol}), (\ref{eqn:tisig}), the following useful formulae can be obtained:
\begin{eqnarray}
\frac{\partial w}{\partial z}=\frac{1}{g(z)} , \\
f_\kappa(z)=e^{\kappa w}
\end{eqnarray}
It is then straightforward to rewrite Eq. (\ref{eqn:cLkapdf}) in terms of $w$:
\begin{equation}
\L_{\kappa}=\frac{1}{2\pi i}\int dw e^{\kappa w}\left. \left(T'(w)+\frac{c}{24}\right) \right|_{t=const.}, \label{eqn:Lkinw}
\end{equation}
where $T'$ represents the energy-momentum tensor in the $w$ coordinate and the term $c/24$ arises from the Schwarzian derivative.

%
%
%
%

\subsection{Expansion of primary fields and the inverse}\label{subsec:Epfi}
The analysis just presented in Sec. \ref{subsec:Va} can be summarised as a clarification of the algebraic structure vis-\`a-vis a specific  (quasi-) primary field $T(z)$ by  integrating over the equal-time path. In this subsection, we try to extend this analysis to involve the general primary fields of CFT.

By definition, a (chiral) primary field of CFT with the conformal dimension $h$ is subject to the following operator-product expansion (OPE):
\begin{equation}
T(z_{1})\phi(z_{2})=\frac{h}{(z_{1}-z_{2})^{2}}\phi(z_{2})+\frac{1}{z_{1}-z_{2}}\frac{\partial \phi(z_{2})}{\partial{z_{2}}}+\dots \ .
\end{equation}
%
%
%
Based on this OPE,  if we integrate $\phi(z)$ along with $g(z)$ and $f_{\kappa}$ as follows:
\begin{equation}
\phi_{\kappa}=\frac{1}{2\pi i}\oint \kern-.6em \ ^{t}\kern.2em dz g^{h-1}(z)f_{\kappa}(z)\phi(z), \label{eqn:phikadf}
\end{equation}
then we can show that the adjoint action of the Virasoro generator $\L_{\kappa}$ on $\phi_{\kappa}$'s amounts to
\begin{equation}
[\L_{\kappa}, \phi_{\kappa'}]=\left( \left(h-1\right)\kappa -\kappa'\right)\phi_{\kappa+\kappa'}. \label{eqn:Lphicom}
\end{equation}
Therefore, by Eq. (\ref{eqn:phikadf}), one can properly define the expansion of a primary field in terms of $\kappa$.
To prove Eq. (\ref{eqn:Lphicom}), simply insert the integral expressions for $\L_{\kappa}$ and $\phi_{\kappa}$, which gives
\begin{eqnarray}
[\L_{\kappa}, \phi_{\kappa'}]=\oint \kern-.6em \ ^{t}\kern.2em \frac{dz_{2}}{2\pi i} && g^{h-1}(z_{2})f_{\kappa'}(z_{2})
\int_{\includegraphics[width=1.5em]{z1aroundz2.pdf}} dz_{1}g(z_{1})f_{\kappa}(z_{1})
\nonumber \\&& \times \left( \frac{h}{(z_{1}-z_{2})^{2}}\phi(z_{2})+\frac{1}{z_{1}-z_{2}}\frac{\partial \phi(z_{2})}{\partial{z_{2}}} \right),
\end{eqnarray}
and similarly with Eq. (\ref{eqn:LLcom}).
The residue integral and the additional partial integral yields
\begin{eqnarray}
\oint \kern-.6em \ ^{t}\kern.2em \frac{dz_{2}}{2\pi i}g^{h-1}(z_{2})f_{\kappa'}(z_{2})
\frac{\partial} {\partial z_{1}} \left(g(z_{1})f_{\kappa'}(z_{1})\right)|_{z_{1}=z_{2}}h \phi(z_{2})\nonumber \\
-\oint \kern-.6em \ ^{t}\kern.2em \frac{dz_{2}}{2\pi i}\frac{\partial}{\partial{z_{2}}}\left(g^{h}(z_{2})f_{\kappa}(z_{2})f_{\kappa'}(z_{2)}\right) \phi(z_{2}).
\end{eqnarray}
With the help of Eq. (\ref{eqn:derfk}), simple arithmetic tells us that the above expression equals
\begin{equation}
\oint \kern-.6em \ ^{t}\kern.2em \frac{dz_{2}}{2\pi i}\left( \left(h-1\right)\kappa -\kappa'\right)g^{h-1}(z_{2})f_{\kappa+\kappa'}(z_{2})\phi(z_{2}),
\end{equation}
which is the right-hand side of Eq. (\ref{eqn:Lphicom}). Note that the above derivation is valid regardless of the form of $g(z)$ or $f_{\kappa}$.

We expect that the above analysis  can be extended to the general case involving (the modes of) two chiral primary fields. However, because demonstrating this requires somewhat meticulous calculations, we defer the effort to a future study. For now, we would rather consider the reciprocal process of what we have just performed. We claim that the following reciprocal expression holds:
\begin{equation}
\phi(z)=g^{-h}(z)\int  d\kappa f_{\kappa}^{-1}(z)\phi_{\kappa}. \label{eqn:phizr}
\end{equation}
Here, the integral over $\kappa$ should be understood as a summation when $\kappa$ takes on discrete values. This relationship is also valid for the quasi-primary energy-momentum tensor $T(z)$ with $h=2$:
\begin{equation}
T(z)=g^{-2}(z)\int  d\kappa f_{\kappa}^{-1}(z)\L_{\kappa}. \label{eqn:Lzre}
\end{equation}

For $g(z)=z$, Eq. (\ref{eqn:phizr}) takes the simple form
\begin{equation}
\phi(z)=\sum_{\kappa} z^{-\kappa -h}\phi_{\kappa},
\end{equation}
which is the well-known Laurent expansion of primary fields. For $g(z)=-(z-1)^{2}/2$, Eq. (\ref{eqn:phizr}) becomes
\begin{equation}
\phi(z)=\frac{1}{\left(-\frac12\left(z-1\right)^{2}\right)^{h}}\int  d\kappa e^{\frac{2\kappa}{z-1}}\phi_{\kappa}. \label{eqn:pzpk}
\end{equation}
The explicit form of Eq. (\ref{eqn:phikadf}) is
\begin{equation}
\phi_{\kappa}=\frac{1}{2\pi i}\oint \kern-.6em \ ^{t}\kern.2em dz' {\left(-\frac12\left(z'-1\right)^{2}\right)}^{h-1}e^{-\frac{2\kappa}{z'-1}}\phi(z'). \ \label{eqn:pkpz}
\end{equation}
Let us take the integral path of $z'$ at $t=0$ and, with the help of Eq. (\ref{eqn:tszdi}), express Eq. (\ref{eqn:pkpz}) in terms of the integral over $s$:
\begin{equation}
\phi_{\kappa}=\frac{1}{2\pi} \int^{-\infty}_{\infty} ds \left(\frac{2}{s^{2}}\right)^{h}e^{-i\kappa s}
\phi\left(1+\frac{2}{is}\right).\end{equation}
Inserting the above into the right-hand side of Eq. (\ref{eqn:pzpk}) yields
\begin{equation}
{\left(-\frac12\left(z-1\right)^{2}\right)}^{-h} \frac{1}{2\pi} \int d\kappa \int^{-\infty}_{\infty} ds e^{i\kappa\left(\frac{2}{i(z-1)}-s\right)}\left(\frac{2}{s^{2}}\right)^{h} \phi\left(1+\frac{2}{is}\right).
\end{equation}
Formally integrating over $\kappa$ yields a delta function; hence,
\begin{equation}
{\left(-\frac12\left(z-1\right)^{2}\right)}^{-h}   \int^{-\infty}_{\infty} ds \left(\frac{2}{s^{2}}\right)^{h} \phi\left(1+\frac{2}{is}\right)\delta\left(\frac{2}{i(z-1)}-s\right).
\end{equation}
Integrating over $s$ yields
\begin{equation}
{\left(-\frac12\left(z-1\right)^{2}\right)}^{-h}   \left(-\frac12(z-1)^{2}\right)^{h} \phi\left(z\right)=\phi(z),\end{equation}
which shows that Eq. (\ref{eqn:phizr}) is valid for  dipolar quantization.

\subsection{Continuum spectrum}\label{seq:contspect}
One immediate consequence of the Virasoro algebra with a continuous index is the continuum spectrum of the system. Consider an eigenstate of $\L_{0}$ with an eigenvalue $\alpha$ and with an additional index $\sigma$ denoting a possible degeneracy:
\begin{equation}
|\alpha , \sigma \rangle  ,
\end{equation}
so that
\begin{equation}
 \L_{0}|\alpha , \sigma \rangle=\alpha |\alpha , \sigma \rangle .
\end{equation}
In this case, operating on $|\alpha , \sigma \rangle$ with $\L_{\kappa}$  yields
\begin{equation}
\L_{\kappa}|\alpha , \sigma \rangle =|\alpha-\kappa , \sigma \rangle .
\end{equation}
This result is based on  the commutation relation (\ref{eqn:Virkap}). Thus, starting from the vacuum or any other energy eigenstate, we can construct an eigenstate for $\L_{0}$ with an arbitrary eigenvalue because $\kappa$ can assume any real value. One way to construct the state $|\alpha, \sigma\rangle$ for non-zero $\alpha$ is to consider
\begin{equation}
\phi_{-\alpha}|0\rangle, \label{pa0}
\end{equation}
where  $\phi_{-\alpha}$ is the expansion of a primary field at $\kappa=-\alpha$. For the additional degeneracy  $\sigma$, one can consider
\begin{equation}
\phi_{0}|0\rangle, \label{p00}
\end{equation}
which is a state with zero energy and is orthogonal to the vacuum, provided
\begin{equation}
\langle 0|\phi_{0}|0\rangle=0.
\end{equation}

%

We now recall that, for the case of ordinary 2d CFTs,  the equation
\begin{equation}
L_{n}|0\rangle=0  \ \  \hbox{for} \  n \geq -1 ,  \label{eqn:Lnv0}
\end{equation}
assures that the energy spectrum of the 2d CFT is bounded below at $|0\rangle$.  
The argument leading to Eq. (\ref{eqn:Lnv0}) is
based on the regularity of the product of the vacuum and the energy-momentum tensor that is placed at $t = -\infty \ (z=0)$, 
\begin{equation}
\lim_{t\rightarrow-\infty}T(z)|0\rangle=\lim_{z\rightarrow0}\sum L_{n} z^{-n-2}|0\rangle. \label{eqn:Txv}
\end{equation}
Were it not for Eq. (\ref{eqn:Lnv0}), Eq. (\ref{eqn:Txv}) would be divergent. Let us examine a suitable variant of (\ref{eqn:Txv}) for the case of dipolar quantization.

The reciprocal expression (\ref{eqn:Lzre}) yields
\begin{equation}
\lim_{t\rightarrow-\infty }T(z)|0\rangle=\lim_{t\rightarrow-\infty }\int_{-\infty}^{\infty}  d\kappa \frac{ (t+is)^{4}}{4}e^{-\kappa(t+is)}\L_{\kappa}|0\rangle. \label{eqn:tmiT0}
\end{equation}
A divergence arises from the factor $e^{-\kappa t}$ because we take $t$ to be $-\infty$ for any positive $\kappa$. Thus, for  Eq. (\ref{eqn:tmiT0}) to be regular at $t\rightarrow-\infty$,  we are led to 
\begin{equation}
\L_{\kappa}|0\rangle=0 \  \hbox{for}\  \kappa >0. \label{eqn:Lkp0}
\end{equation}
If we apply the same argument to a primary field $\phi(z)$ with the conformal weight $h$, and note that
\begin{equation}
\lim_{t\rightarrow-\infty }\phi(z)|0\rangle=\lim_{t\rightarrow-\infty }\int_{-\infty}^{\infty}  d\kappa \frac{ (t+is)^{2h}}{(-2)^{h}}e^{-\kappa(t+is)}\phi_{\kappa}|0\rangle, \label{eqn:tipzpk}
\end{equation}
we also must demand
\begin{equation}
\phi_{\kappa}|0\rangle=0 \  \hbox{for}\  \kappa >0. \label{eqn:phkp0}
\end{equation}
However,  the exponential factor $e^{-\kappa t}$ suppresses any contribution from $\kappa <0$ in Eq. (\ref{eqn:tipzpk}), which suggests
\begin{equation}
\lim_{t\rightarrow-\infty }\phi(z)|0\rangle=\phi(1)|0\rangle=\phi_{0}|0\rangle .
\end{equation}
This expression  should be compared with that for the highest-weight state for radial quantization case:
\begin{equation}
\phi(0)|0\rangle=\phi_{-h}|0\rangle.
\end{equation}

Equation (\ref{eqn:Lkp0}) implies that the spectrum of the system is bounded below by $|0\rangle$, at least for the states derived from the multiplication of $\L_{\kappa}$, starting from either the vacuum $|0\rangle$ or the states (\ref{pa0}). Therefore, the Hamiltonian
\begin{equation}
H=\L_{0}+{\bar \L}_{0}
\end{equation}
possesses a continuous spectrum. The expectation expressed in the introduction is thus verified. This result is also consistent with the observation that the variable $s$ takes values from $-\infty$ to $\infty$.

The general structure of the Hilbert space for  dipolar quantization is expected to be complex because it has a continuous index. In particular, in the above construction, there are many ways to multiply $L_{\kappa}$ to obtain a given eigenvalue, implying that each eigenstate is heavily degenerate. Although a complete analysis of the Hilbert space is beyond the scope of this paper, we do discuss some aspects of the Hermitian conjugate to clarify the structure of the Hilbert space in Sec. \ref{sec:HCHS}.

\section{Hermitian conjugate and  Hilbert space}\label{sec:HCHS}
In the previous section, we  considered a set of vectors such as
\begin{equation}
|\kappa\rangle_{0} \equiv \L_{-\kappa}|0\rangle, \kappa >0, \label{eqn:Lmk0vs}
\end{equation}
which provide the basis of the continuous spectrum for the dipolar-quantized system. It would be desirable if they form an orthonormal set of vectors. In fact,
if we assume the Hermitian conjugate takes the following form for  dipolar quantization:
\begin{equation}
\L_{\kappa}^{\dagger}=\L_{-\kappa}, \label{eqn:HmtcnjL}
\end{equation}
the inner product of the vectors mentioned above,
\begin{equation}
{}_{0}\langle \kappa' |\kappa\rangle_{0} =\langle 0| \L_{\kappa'}\L_{-\kappa}|0\rangle=\langle 0|[ \L_{\kappa'},\L_{-\kappa}]|0\rangle,
\end{equation}
amounts  to 
\begin{equation}
{}_{0}\langle \kappa' |\kappa\rangle_{0} =\frac{c}{12}\kappa^{3}\delta(\kappa'-\kappa)\langle 0 |0\rangle, \label{eqn:nbvs}
\end{equation}
owing to Eq. (\ref{eqn:Virkap}). Therefore, the above assumption (\ref{eqn:HmtcnjL}) seems natural and also implies  the Hermiticity of $\L_{0}$. We justify  the assumption (\ref{eqn:HmtcnjL})  in the following.

First, let us reflect upon  Hermitian conjugation
%
\footnote{By slight abuse of terminology,  {\it Hermitian conjugation} ($\dagger$) that we refer here is actually [conjugation]$\times$[time-reversal]  when the operator depends on Euclidean time. In Ref. \cite{Polchinski:1998rq}, this is termed {\it Euclidean adjoint}  in order to avoid confusion.} 
in general.
The analysis of Hermitian conjugation is simpler in terms of $t$ and $s$ coordinates rather than that of $z$.
Because we are considering a field theory in  Euclidean space, the time coordinate is actually  imaginary:
\begin{equation}
t=i\tau, \label{eqn:titau}
\end{equation}
where $\tau$ is Minkowski time. This extra $i$ stems from the difference between the time development of an operator in the Euclidean time, $\phi(t)=e^{Ht}\phi(0)e^{-Ht}$, and that in the Minkowski time, $\phi(\tau)=e^{iH\tau}\phi(0)e^{-iH\tau}$.
Thus, when applying the Hermitian conjugate in terms of the corresponding Hilbert space  to a field operator, the imaginary unit $i$ in Eq. (\ref{eqn:titau}) should be  properly taken into account.


As just mentioned, the operator for arbitrary $\tau$ is derived from the time translation generated by the Hamiltonian $H$:
\begin{equation}
\phi(\tau)=e^{iH\tau}\phi(0)e^{-iH\tau},
\end{equation}
where $H$ is assumed to be Hermitian.
In addition,  we are only concerned with  field operators that are real fields in Minkowski space. This means that  each operator on a certain spacetime point  is  Hermitian: 
\begin{equation}
\left(\phi(0)\right)^\dagger=\phi(0).
\end{equation}
Then, the corresponding Heisenberg operator for $\tau$ is also Hermitian:
\begin{equation}
\left(\phi(\tau)\right)^\dagger=\left(e^{iH\tau}\phi(0)e^{-iH\tau}\right)^\dagger=e^{iH\tau}\phi(0)e^{-iH\tau}=\phi(\tau).
\end{equation}
However, if we repeat the above procedure in  Euclidean time,
the Heisenberg operator is no longer Hermitian; since,
\begin{equation}
\left(\phi(t)\right)^\dagger=\left(e^{Ht}\phi(0)e^{-Ht}\right)^\dagger=e^{-Ht}\phi(0)e^{Ht}=\phi(-t).
\end{equation}
In order for the Heisenberg operator $\phi(t)$ to be Hermitian, Hermitian conjugation  must be accompanied by reversal of Euclidean time $t$:
\begin{equation}
t\mapsto -t . \label{eqn:tmt}
\end{equation}

%
%
%
%
%


There is another way to understand the requirement of time-reversal for Euclidean time. 
For the time-development to be unitary, the following must hold:
\begin{equation}
\left(e^{Ht}\right)^\dagger=e^{-Ht},
\end{equation}
which shows that  Hermitian conjugation has to be accompanied with the change of the sign of $t$, $t\rightarrow-t$, otherwise  the time-development operator $e^{Ht}$ would not  be unitary.
In terms of the complex coordinate $w$
\begin{equation}
w=t+is, \label{eqn:wdef}
\end{equation}
the change in the sign of $t$ [see Eq. (\ref{eqn:tmt})] is understood as
\begin{equation}
w\mapsto -{\bar w}.
\end{equation}

Therefore, Hermitian conjugation eventuate in  complex conjugation of the variable, which is rather desirable because complex conjugation is usually associated with Hermitian conjugation, and ${\bar z}$ is  expected to emerge upon Hermitian conjugation on $z$ plane.

\begin{figure}[tbh]
\centering
\includegraphics[width=15cm]{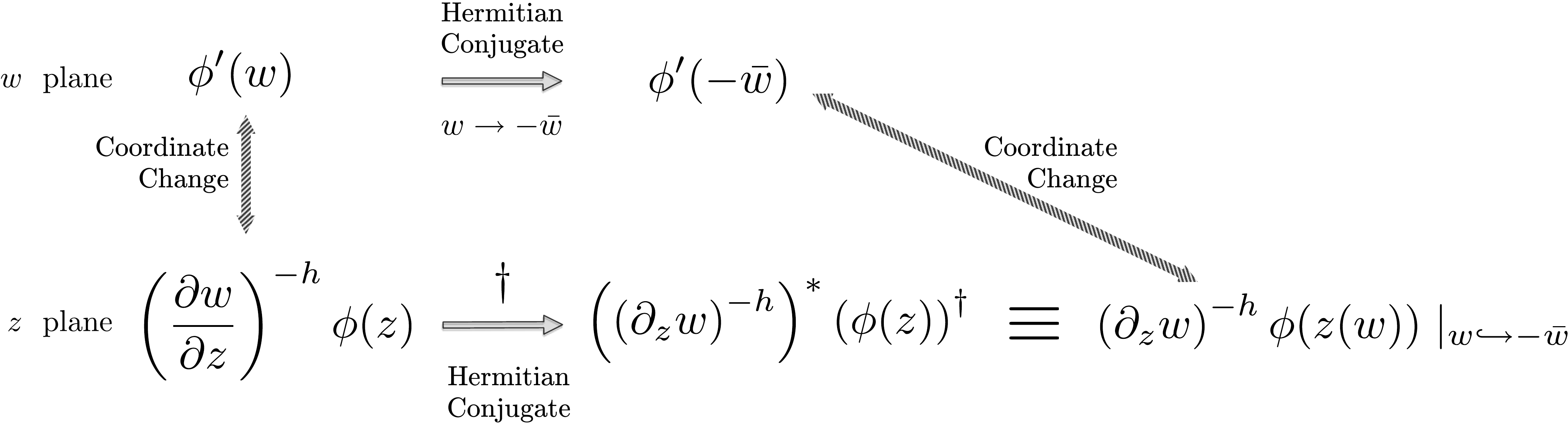}
\caption{Illustration of definition of Hermitian conjugation on $z$ plane. The simplicity of the procedure on $w$ plane can be exploited to transport it to the $z$ plane by way of the coordinate change. The notion $^{*}$ denotes  complex conjugation.
} \label{fig:hmtcnj}
\end{figure}

We now exploit the simplicity of the above definition of  Hermitian conjugation on the $w$ plane and apply it to the $z$ plane.
Putting aside the Schwarzian derivative term for the case of the energy-momentum tensor,
the change of the coordinates between $z$ and $w$ $(z\leftrightarrow w)$ evokes the following relationship for the operator with conformal weight $h$:
\begin{equation}
\phi'(w)=\left(\frac{\partial w}{\partial z}\right)^{-h}\phi(z). \label{eqn:pdwjpz}
\end{equation}
For  ordinary radial quantization,
\begin{equation}
z=e^{w} \ , \ \  w=\ln z,
\end{equation}
and Eq. (\ref{eqn:pdwjpz}) becomes
\begin{equation}
\phi'(w)=z^{h}\phi(z). \label{eqn:pdwzhpz}
\end{equation}
Taking the Hermitian conjugate on the $w$ plane yields
\begin{equation}
\left( \phi'(w) \right)^{\dagger} = \phi'(-{\bar w}). \label{eqn:pdwhmt}
\end{equation}
By using Eq. (\ref{eqn:pdwzhpz}), the left-hand side of Eq. (\ref{eqn:pdwhmt}) becomes
\begin{equation}
\left( \phi'(w) \right)^{\dagger} =\left( z^{h}\phi(z) \right)^{\dagger} =(\bar z)^{h}\left(\phi(z)\right)^{\dagger}.\label{eqn:pdwhmt1}
\end{equation}

The right-hand side of Eq. (\ref{eqn:pdwhmt}) can be estimated by replacing $w$ with $-{\bar w}$ in Eq. (\ref{eqn:pdwzhpz}):
\begin{equation}
\phi'(-{\bar w})=e^{-h{\bar w}}\phi\left(e^{-{\bar w}}\right)={\bar z}^{-h}\phi\left(\frac{1}{\bar z}\right).\label{eqn:pdwhmt2}
\end{equation}
Equating the results of Eqs. (\ref{eqn:pdwhmt1}) and (\ref{eqn:pdwhmt2}) leads to the natural definition of Hermitian conjugation that affirms the unitary time-development in Minkowski space-time:
\begin{equation}
\left(\phi(z)\right)^{\dagger}=({\bar z})^{-2h}\phi\left(\frac{1}{\bar z}\right). \label{eqn:rdHmtc}
\end{equation}
The argument presented above is summarised in Fig. \ref{fig:hmtcnj}.
Equation (\ref{eqn:rdHmtc}) holds for the energy-momentum tensor $T(z)$ with $h=2$, because the extra Schwarzian derivative that would appear in Eq. (\ref{eqn:pdwjpz})  cancels out in the subsequent comparison.

The above argument can be repeated for dipolar quantization.
From Eq. (\ref{eqn:tszdi}),
\begin{equation}
z=1+\frac{2}{w}\ , \ \ w=\frac{2}{z-1},
\end{equation}
and
\begin{equation}
\frac{\partial w}{\partial z}=-\frac{2}{(z-1)^{2}}=-\frac{w^{2}}{2}.
\end{equation}
Equation (\ref{eqn:pdwjpz}) can be written  explicitly as
\begin{equation}
\phi'(w)=\left(-\frac{2}{(z-1)^{2}}\right)^{-h}\phi(z)=\left(-\frac{w^{2}}{2}\right)^{-h}\phi\left(1+\frac{2}{w}\right).
\end{equation}
The left-hand side of Eq. (\ref{eqn:pdwhmt}) for  dipolar quantization then becomes
\begin{equation}
\left( \phi'(w) \right)^{\dagger} = \left( \left(-\frac{2}{(z-1)^{2}}\right)^{-h}\phi(z)\right)^{\dagger}= \left(-\frac{2}{({\bar z}-1)^{2}}\right)^{-h}\left(\phi(z)\right)^{\dagger}, \label{eqn:pdwhmt3}
\end{equation}
and the right-hand side of Eq. (\ref{eqn:pdwhmt}) is
\begin{equation}
\phi'(-{\bar w})=\left(-\frac{(-{\bar w})^{2}}{2}\right)^{-h}\phi\left(1-\frac{2}{{\bar w}}\right) = \left(-\frac{2}{({\bar z}-1)^{2}}\right)^{-h}\phi(2-{\bar z}).\label{eqn:pdwhmt4}
\end{equation}
Comparing Eqs. (\ref{eqn:pdwhmt3}) and (\ref{eqn:pdwhmt4}), we conclude that
\begin{equation}
\left(\phi(z)\right)^{\dagger}=\phi(2-{\bar z}),\label{eqn:diHmtc}
\end{equation}
for  dipolar quantization. In particular, if applied to the energy-momentum tensor,
\begin{equation}
\left(T(z)\right)^{\dagger}=T(2-{\bar z}).\label{eqn:TdiHmtc}
\end{equation}

Let us now reflect on the above result based on physical intuition. One of the physical implications of Hermitian conjugation on the Hilbert space in which we construct quantum theory is the exchange of the states at $t\rightarrow -\infty$ and $t\rightarrow \infty$. For  radial quantization, the state for $t\rightarrow -\infty$ is located at $z=0$, whereas $z\rightarrow \infty$ accommodates the state for $t\rightarrow \infty$. This result is consistent with the expression (\ref{eqn:rdHmtc}), because it relates the variables at $z=0$ and $\infty$ by way of $1/{\bar z}$. Now, for  dipolar quantization, both $t\rightarrow -\infty$ and $t\rightarrow \infty$ drive a state to the same point $z=1$. Putting $z=1$ in Eq. (\ref{eqn:diHmtc}) gives $\left(\phi(1)\right)^{\dagger}=\phi(1)$, which confirms the validity of the Hermitian conjugation derived above.

Based on Eq. (\ref{eqn:cLkapdf}), the Virasoro generators for radial quantization are written  as
\begin{equation}
\L_{n} = L_{n}=\frac{1}{2\pi i}\oint   dz z^{n+1}T(z), 
\end{equation}
and the Hermitian conjugate is
\begin{equation}
L_{n}^{\dagger} = -\frac{1}{2\pi i}\oint   d{\bar z} {\bar z}^{n+1}\left(T(z)\right)^{\dagger}=-\frac{1}{2\pi i}\oint  d{\bar z} {\bar z}^{n+1}{\bar z}^{-4}T\left(\frac{1}{\bar z}\right).
\end{equation}
Rewriting ${\bar z}$ with a new variable $z'$ and considering the direction of the contour with care yields
\begin{equation}
L_{n}^{\dagger}= \frac{1}{2\pi i}\oint  d{z'} {(z')}^{n-3}T\left(\frac{1}{ z' }\right).
\end{equation}
By further changing the variables from $z'$ to $z''=1/z'$, we arrive at
\begin{equation}
L_{n}^{\dagger}= \frac{1}{2\pi i}\oint  d{z''} {(z'')}^{-n+1}T(z'')=\L_{-n} =L_{-n}.
\end{equation}
The above result simply confirms the well-known fact.

\begin{figure}[hbt]
\centering
\includegraphics[width=12cm]{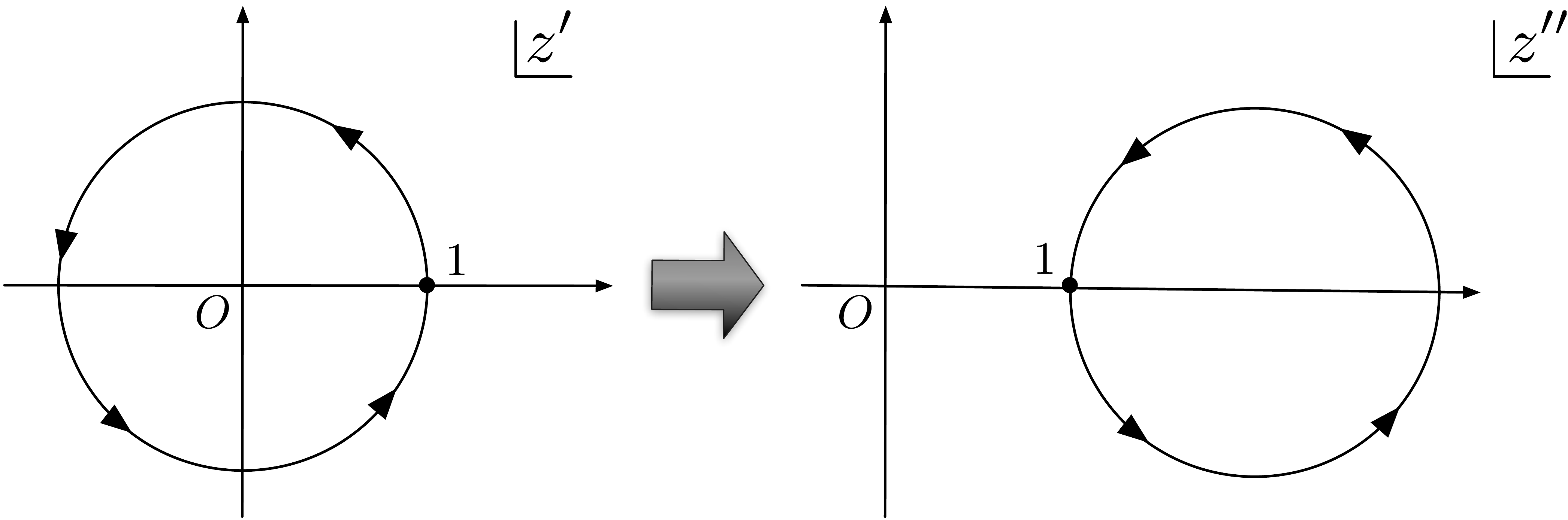}
\caption{Contour is altered by the change of the variables from $z'$ to $z''=2-z'$. Comparing the contour on the right with the contour depicted in Fig. \ref{fig:equaltimepath2} $(c)$ explains how the additional negative sign appears.} \label{fig:2mz}
\end{figure}
The point here is that exactly the same argument applies to dipolar quantization. In this case, the Virasoro generators are
\begin{equation}
\L_{\kappa}=\int_{\includegraphics[width=1.5em]{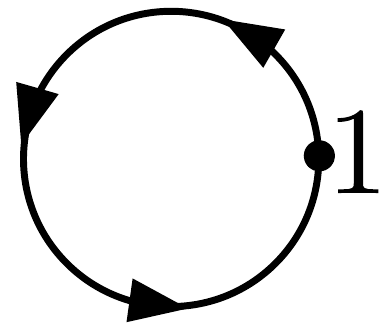}} dz\left(-\frac12\left(z-1\right)^{2}\right)e^{\frac{2\kappa}{z-1}}T(z),
\end{equation}
where we take an integration path such that time $t$ is  negative; for example, $t=-1$ (see Fig. \ref{fig:equaltimepath2}).
The Hermitian conjugate is
\begin{eqnarray}
\L_{\kappa}^{\dagger} &=& -\frac{1}{2\pi i}\int_{\includegraphics[width=1.5em]{eqtconz.pdf}}   d{\bar z} \left(-\frac12\left({\bar z}-1\right)^{2}\right)e^{\frac{2\kappa}{{\bar z}-1}}\left(T(z)\right)^{\dagger} \nonumber \\
&=&-\frac{1}{2\pi i}\int_{\includegraphics[width=1.5em]{eqtconz.pdf}}     d{\bar z} \left(-\frac12\left({\bar z}-1\right)^{2}\right)e^{\frac{2\kappa}{{\bar z}-1}}T\left(2-{\bar z}\right). \label{eqn:Lkdbz}
\end{eqnarray}
Rewriting ${\bar z}$ with a new variable $z'$ yields
\begin{equation}
\L_{\kappa}^{\dagger}=\frac{1}{2\pi i}\int_{\includegraphics[width=1.5em]{eqtconz.pdf}}     d{z'} \left(-\frac12\left({z'}-1\right)^{2}\right)e^{\frac{2\kappa}{{z'}-1}}T\left(2-{z'}\right),
\end{equation}
where the direction of the contour was inverted once because of the change of variables from an anti-holomorphic variable to a holomorphic variable, but  the absorption of the extra $-1$ that appeared in front of the integral in Eq. (\ref{eqn:Lkdbz}) inverted the contour back to the original direction.
By further changing  variables  from $z'$ to $z''=2-z'$, we arrive at
\begin{equation}
\L_{\kappa}^{\dagger}=\frac{1}{2\pi i}\int_{\includegraphics[width=1.5em]{eqtconz.pdf}}     d{z''} \left(-\frac12\left({z''}-1\right)^{2}\right)e^{-\frac{2\kappa}{{z''}-1}}T\left({z''}\right)=\L_{-\kappa}. \label{eqn:dLkdLmk}
\end{equation}
In the above derivation leading to Eq. (\ref{eqn:dLkdLmk}), the negative sign from the change of variables was cancelled by another negative sign from the change in the direction of the contour, as explained in Fig. \ref{fig:2mz}.

Thus, we have established Hermitian conjugation in the form of Eq. (\ref{eqn:HmtcnjL}) and the subsequent expression of orthogonality (\ref{eqn:nbvs}) among the vectors, at least in the form of Eq. (\ref{eqn:Lmk0vs}).

In the framework of  dipolar quantization, we can still consider the following operators as the contour integrals of the energy-momentum tensor
over the path where time is set to  $-1$:
\begin{equation}
\frac{1}{2\pi i}\int_{\includegraphics[width=1.5em]{eqtconz.pdf}}    dz z^{n+1}T(z).
\end{equation}
This expression is apparently identical to the definition of $L_{n}$ in radial quantization. We also denote this expression as $L_{n}$ and apply Hermitian conjugation as defined above, which yields
\begin{equation}
(L_{n})^{\dagger}=\frac{1}{2\pi i}\int_{\includegraphics[width=1.5em]{eqtconz.pdf}}     d{z''} \left(2-z''\right)^{n+1}T\left({z''}\right). \label{eqn:dLnHc}
\end{equation}
The cases of $n=-1, 0, 1$ produce a particularly interesting result. From Eq. (\ref{eqn:dLnHc}),
\begin{eqnarray}
(L_{-1})^{\dagger}&=&L_{-1} ,\nonumber \\
(L_{0})^{\dagger}&=&2L_{-1} -L_{0},\label{eqn:L-101}\\
(L_{1})^{\dagger}&=&L_{1}-4L_{0}+4L_{-1}. \nonumber 
\end{eqnarray}
Equation (\ref{eqn:L-101}) shows that the operations of the Hermitian conjugation operations in  dipolar quantization on $L_{-1}, L_{0}$ and $L_{1}$ are closed among themselves. In addition, they definitely take a different form from those of  radial quantization. Nonetheless, if we compute the Hermitian conjugate for the combination $L_{0}-(L_{1}+L_{-1})/2$, which is the Hamiltonian for dipolar quantization, it proves to be Hermitian (in the sense of dipolar quantization):
\begin{eqnarray}
\left(L_{0}-\frac{L_{1}+L_{-1}}{2}\right)^{\dagger}&=&2L_{-1} -L_{0}-\frac{L_{1}-4L_{0}+4L_{-1}+L_{-1} }{2}\nonumber \\
&=&L_{0}-\frac{L_{1}+L_{-1}}{2}.
\end{eqnarray}

Since the definition of the Hermitian conjugation is straightforward in the $w$ coordinate, one can also prove  Eq. (\ref{eqn:HmtcnjL}) by using Eq. (\ref{eqn:Lkinw}), which has a particularly simple expression if we choose $t=0$:
\begin{equation}
\L_{\kappa}=\frac{1}{2\pi}\int ds e^{i\kappa s}\left(T'(is)+\frac{c}{24}\right). \label{eqn:Lkis}
\end{equation}
In the $w$ coordinate, the effect of the Hermitian conjugation is  reversal of Euclidean time as in Eq. (\ref{eqn:tmt}); however, setting $t=0$ negates this effect. Therefore, applying the Hermitian conjugation on Eq. (\ref{eqn:Lkis}) yields
\begin{equation}
\L_{\kappa}^{\dagger}=\frac{1}{2\pi}\int ds e^{-i\kappa s}\left(T'(is)+\frac{c}{24}\right)=\L_{-\kappa}.
\end{equation}

\section{Summary and Discussion}\label{sec:SMDI}
We have shown in this report that 2d CFT admits an alternative quantization other than usual radial quantization and that we call dipolar quantization. Although the conformal symmetry quite determines the nature of the energy-momentum tensor and other (primary) operators,  choosing different time-translations remains possible. This possibility is suggested by  consideration of the $sl(2, \mathbb{R})$ subalgebra of the conformal symmetry. 

One salient result  is that the different quantization procedures yield different spacetime: finite space for radial quantization, and infinite space for dipolar quantization. Usually the study of the quantum field theory starts by defining the spacetime where the field is situated. Here, we first obtain the quantum system, and its analysis  reveals the nature of spacetime. In this sense, the quantum system may be considered more fundamental than the classical notion of spacetime. This viewpoint is in accordance with the efforts to consider spacetime as emergent phenomenon from the fundamental quantum system; for example, matrix models
\cite{Aoki:1998vn}.

Although we  reveal several key facts, such as the difference of the Hilbert space structure, the emergence of the continuous spectrum, and the vacuum degeneracy (\ref{p00}), this study  is  limited and  many aspects remain to be clarified. Nonetheless, these aforementioned facts might provide sufficient motivation to revisit some basic features of CFT in the context of dipolar quantization; for example, the Zamoldchikov-Polchinski Theorem \cite{Zamolodchikov:1986gt,Polchinski:1987dy}.

As just mentioned, $SL(2, \mathbb{R}) $ symmetry plays a crucial role in our analysis.  
In fact, $SL(2, \mathbb{R}) $ or $SL(2, \mathbb{C}) \sim SL(2, \mathbb{R})\otimes SL(2, \mathbb{R}) $ invariance holds for CFT in any dimension. Therefore, one may be tempted to imagine that for dimensions other than two, the the procedure of dipolar quantization can also be employed with suitable differential operators such as ${\hat D}-({\hat K}_{\mu}+P_{\mu})/2$ .

Another interesting aspect to be explored is supersymmetry. Applying dipolar quantization to  2d superconformal field theory (SCFT) turns out to be rather straightforward, as summarised in Ref.\ref{sec:SCFT}. However, it is conceivable that supersymmetry may play much more important roles in future study of  dipolar quantization. In fact, a recent finding \cite{Okunishi:2015dfa} reveals an intriguing relationship between supersymmetry and SSD. It would be interesting to see if this is related to the supersymmetric representation of $SL(2, \mathbb{R})$ \cite{Ramond:2010zz}.

The present formulation was also partially  guided by  previous approaches in the study of string field theory (SFT) \cite{Kiermaier:2007jg, Takahashi:2003xe}. It would be interesting if one can find more direct connections between the present result and SFT treatmen, especially in the context of understanding the transition between open and closed strings \cite{Takahashi:2011zzb}.

We have emphasised the distinction between the present formulation and the tensionless string in the introduction. However, it would still be interesting to explore a connection or physical implication from the study of long strings  \cite{Maldacena:2005hi,Seiberg:2005nk,Kostov:2006dy}.

Finally, one of the present authors encountered a severe divergence in \cite{Tada:2014kza} when he tried to find the Lagrangian that corresponds to the SSD Hamiltonian. The analysis presented in this report suggests that the  divergence stems from the apparent innocuous assumption that the Lagrangian can be obtained from the integration over the finite space. It would be interesting to see if  the Lagrangian can be constructed on an infinitely large space that corresponds to the current formulation.

\vspace{1em}\noindent{\bf Acknowledgements:} 
This study is supported in part by JSPS KAKENHI Grant No. 25400242, JSPS KAKENHI Grant No. 25610066 and the RIKEN iTHES Project. We would like to thank M. Asano,  H. Itoyama, H. Katsura, H. Kawai, V. Kazakov, I. Kostov, Y. Matsuo, K. Okunishi, N. Sakai, T. Yoneya and S. Zeze  for valuable discussions, comments and inputs. T. T.  is indebted to the Yukawa Institute for Theoretical Physics at Kyoto University for its hospitality during the workshop YITP-W-14-4 `Strings and Fields', where part of the present work was conducted. T. T. would also like to thank D. Lowe for bringing his attention to Ref. \cite{Wybourne:1974}. 

\appendix
\section{Holomorphic transformation of differential operator}\label{sec:HTDO}
In this appendix, we note some effects of a holomorphic transformation on a class of differential operators.
In particular, we consider the following type of the differential operator:
\begin{equation}
p(z)\frac{\partial}{\partial z},
\end{equation}
where $p(z)$ is assumed to be a polynomial in $z$. The holomorphic transformation in which we are interested in is $z'(z)$, such that
\begin{equation}
-p(z)\frac{\partial}{\partial z}=-z'\frac{\partial}{\partial z'}.
\end{equation}
The above condition can be written as
\begin{equation}
-p(z)\frac{\partial}{\partial z}=\frac{\partial z'}{\partial z}\frac{\partial}{\partial z'}=-z'\frac{\partial}{\partial z'};
\end{equation}
hence, we arrive at
\begin{equation}
p(z)\frac{dz'}{dz}=z',
\end{equation}
or
\begin{equation}
\frac{dz'}{z'}=\frac{dz}{p(z)},
\end{equation}
which can be easily integrated to obtain
\begin{equation}
z'(z)=\exp (\int^{z} \frac{dz}{p(z)}).
\end{equation}

For the sake of simplicity, we  consider $p(z)$ in the following form by introducing the parameter $\alpha$:
\begin{equation}
p(z)=z-\alpha(z^{2}+1). \label{eqn:pzalpha}
\end{equation}
Taking $\alpha =0$ yields a trivial transformation, whereas $\alpha=\frac12$ corresponds to the case we have investigated with dipolar quantization. It would be thus suffice to consider  $\alpha >0$. The integration formulas for the inverse of a quadratic polynomial can be summarised in the following compact form \cite{GradshteynRyzhik}:
\begin{equation}
\int \frac{dx}{A+2Bx+C x^{2}}=\left\{\begin{array}{cc}AC >B^{2} & \frac{1}{\sqrt{AC-B^{2}}}
\arctan \frac{Cx+B}{\sqrt{AC-B^{2}}}\\ &  \\AC=B^{2}& -\frac{1}{C\left( x+\frac{B|C|}{|B|C}\sqrt{\frac{A}{C}}\right)} \\  &   \\ AC<B^{2} &  \frac{1}{2\sqrt{B^{2}-AC}}
\ln \left| \frac{Cx+B-\sqrt{B^{2}-AC}}{Cx+B+\sqrt{B^{2}-AC}}\right|\end{array}\right. .
\end{equation}
Using the above formula, we obtain
\begin{equation}
\int \frac{dz}{-\alpha z^{2}+z-\alpha}=\left\{\begin{array}{cc}\alpha > \frac12 & 
\frac{2}{\sqrt{4\alpha^{2}-1}}
\arctan \frac{2\alpha z-1}{\sqrt{4\alpha^{2}-1}}\\ &  \\\alpha=\frac12 & \frac{2}{z-1} \\  &   \\ 0<\alpha < \frac12 &  \frac{1}{\sqrt{1-4\alpha^{2}}}
\ln \frac{2\alpha z+\sqrt{1-4\alpha^{2}}-1}{2\alpha z-\sqrt{1-4\alpha^{2}}-1
}  \end{array}\right. .
\end{equation}
For $0<\alpha<\frac12$, we drop the absolute-value restriction $| \cdot |$ in the logarithm.
The idiosyncrasy of the case $\alpha=\frac12$ is apparent in the above result. In addition, for this case, the holomorphic transformation takes the following particularly simple form:
\begin{equation}
z'(z)=\exp\left(\frac{2}{z-1}\right). \label{eqn:wzdipolar}
\end{equation}
Note that this transformation is the same as that in Eq. (\ref{eqn:wez}).
With this $z'$ coordinate, the usual correspondence between the $z$ and $w$ coordinates holds:
\begin{equation}
w=\ln z'=\frac{2}{z-1}.
\end{equation}

Also quadratic but with a different parametrization, consider
\begin{equation}
p(z)=ac z^{2}+(ad+bc)z +bd\ , \label{eqn:pzabcd}
\end{equation}
where $ad-bc$ is set to unity. This integrates to yield
\begin{equation}
\int \frac{z}{p(z)}=\ln
\frac{az+b}{cz+d}
+\ln
\frac{c}{d}
.
\end{equation}
By omitting the constant part in the above equation, we arrive at
\begin{equation}
z'(z)=\frac{az+b}{cz+d} \ ,
\end{equation}
which is the $sl(2, \mathbb{R})$ transformation on the worldsheet. Note that the two parametrizations above are totally incompatible: if we equate the coefficient $ad+bc$ of $z$ in Eq. (\ref{eqn:pzabcd}) with that of Eq. (\ref{eqn:pzalpha}), which is unity, this contradicts the precondition $ad-bc=1$.

\section{$\mathbf{Z}_{n}$-symmetric equal-time contours}\label{sec:ZnETC}
In this appendix, we explore another set of (differential) operators that generate interesting `time-translations'. First, note that the following generators also form $sl(2, \mathbb{R})$:
\begin{eqnarray}
\left[\frac{l_{n}}{n},\frac{l_{0}}{n}\right]&=&\frac{l_{n}}{n} ,\\
\left[\frac{l_{0}}{n},\frac{l_{-n}}{n}\right]&=&\frac{l_{-n}}{n} ,\\
\left[\frac{l_{n}}{n},\frac{l_{-n}}{n}\right]&=&2\frac{l_{0}}{n} ,
\end{eqnarray}
where $n$ is an integer larger than unity.
If we replace $\{ l_{0}/n, l_{n}/n, l_{-n}/n \}$ with $\{ l_{0}, l_{1}, l_{-1} \}$ in the analysis that leads to Eq. (\ref{eqn:ssdgz}), we obtain
\begin{equation}
g(z)=\frac{1}{n}\left(z-\frac{z^{n+1}+z^{-n+1}}{2}\right)=-\frac{1}{2n}\frac{(z^{n}-1)^{2}}{z^{n-1}},
\end{equation}
in stead. Repeating the analysis of Sec. \ref{sec:generalization}, the expression for the worldsheet parameters $t$ and $s$ can also be obtained by a simple integration as
\begin{equation}
t+is=\int_{z} \frac{dz}{g(z)}=\frac{2}{z^{n}-1}.
\end{equation}
Using the polar coordinates $z=re^{i\theta}$ is convenient for the following analysis:
\begin{eqnarray}
t&=&\frac{2r^{n}\cos n\theta -2}{r^{2n}-2r^{n}\cos n\theta+1} \ ,\label{eqn:tLn}\\
s&=&\frac{-2\sin n\theta}{r^{n}+r^{-n}-2\cos n\theta} \ \label{eqn:sLn} .
\end{eqnarray}

\begin{figure}[tbh]
\centering
\includegraphics[width=16.5cm]{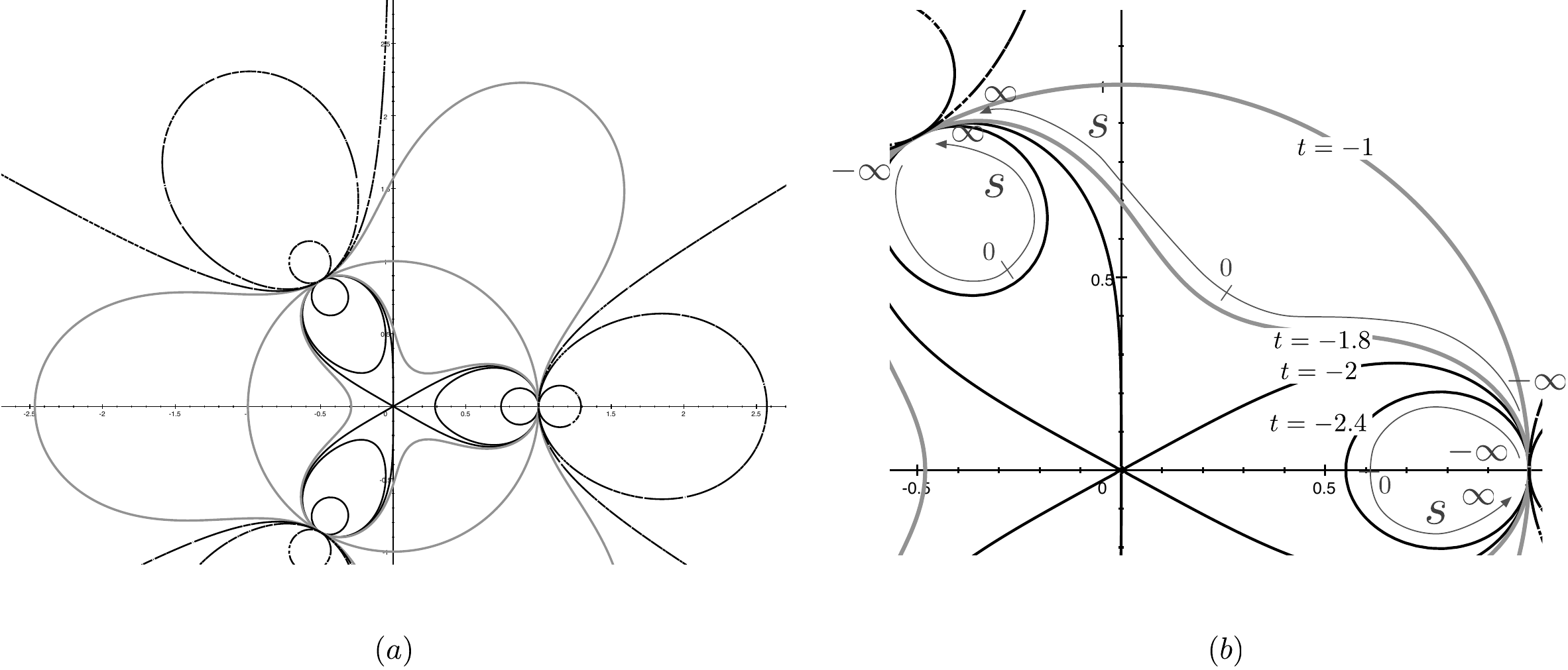}
\caption{Equal-time contours for various values of $t$ are depicted for the $n=3$.  The contour starts to grow towards the inside of the unit circle from three points in the $\mathbf{Z}_{3}$-symmetric way. The contour eventually reaches the origin and forms a single connected contour for $-2<t<0$ (grey lines). For $t>0$, the contour is again divided into three disconnected parts (broken lines).  $(a)$ $\mathbf{Z}_{3}$ symmetry is apparent. $(b)$ The parameter $s$ covers $\mathbb{R}$ three times for both connected and disconnected contours.
} \label{fig:multipolemonochrome}
\end{figure}
%
%
We focus now on some special values of  $\theta$ by introducing $\theta_{m} \sim 0$ as follows:
\begin{equation}
\theta=\frac{2\pi}{n}m+\theta_{m} \ , \ m=0, \dots, n-1 \ \ ,
\end{equation}
The parameter $s$ can then be expanded in terms of $\theta_{m}$ as
\begin{equation}
s \sim -\frac{2n\theta_{m}}{r^{n}+r^{-n}-2+n^{2}\theta_{m}^{2}}.
\end{equation}
This expression yields simply
\begin{equation}
\theta_{m} \rightarrow 0 \Rightarrow s\rightarrow 0
\end{equation}
unless $r =1$.
At $r=1$, 
\begin{equation}
\theta_{m} \rightarrow \pm0 \Rightarrow s \sim -\frac{2}{n\theta_{m}}\rightarrow \mp \infty.
\end{equation}
This indicates that the idiosyncratic role played by the point $z=1$ is also adopted by the other $n-1$ points on the unit circle:
\begin{equation}
z=e^{2\pi \frac{m}{n}} \ , \ m=1, \dots, n-1 \ .
\end{equation}

Thus, as $t\rightarrow -\infty$, the contour converges into $n$ points on the unit circle $z=e^{2\pi \frac{m}{n}}$. Each separated contours then grows in a $\mathbf{Z}_{n}$-symmetric way, as can seen in Fig.\ref{fig:multipolemonochrome}. At $t=-2$,  the contours reach the origin $z=0$ and all the separated parts of the contour  connects at the origin. In fact, if we set $r=0$ in Eq. (\ref{eqn:tLn}), we obtain $t=-2$. Another interesting value for $r$ is  unity. For which Eq. (\ref{eqn:tLn}) gives $t=-1$,  the unit circle. For $t>0$, the contour again separates into $n$ parts.

Note also that $s$ goes to zero not only for $\theta=2 \pi \frac{m}{n}$ but also for  $\theta= \pi \frac{m}{n}$ . Thus, for the time $-2<t<0$,  each segment of the contour connects smoothly at $\theta= 2\pi \frac{m-1}{n}$, where $s=0$, whereas at $\theta=2 \pi \frac{m}{n}$, $s$ goes to either positive or negative infinity depending on the direction of the approach. For the connected contour, $s$ also covers all real numbers $m$ times, which also  trivially holds for the  contour disconnected into $m$ parts.

\section{Superconformal field theory}\label{sec:SCFT}
Within the realm of the analysis presented in the main text, there are interesting additions that extend the structure of conformal field theory: the superconformal field theories. These add extra fermionic current $G(z)$ for $N=1$  and fermionic currents $G^{\pm}(z) $ and $U(1)$ current $J(z)$ for $N=2$. Applying the present analysis explicitly to these special cases is of interest. For this purpose, we derive the relevant algebra in terms of our analysis by assuming the OPE's.

The OPE for $N=1$ SCFT consists of the following set of equations:
\begin{eqnarray}
T(z_{1})T(z_{2})& \sim &\frac{\frac12 c}{(z_{1}-z_{2})^{4}} + \frac{2T(z_{2})}{(z_{1}-z_{2})^{2}}+\frac{\partial_{z_{2}}T(z_{2})}{z_{1}-z_{2}}, \label{eqn:opeN1TT}\\
T(z_{1})G(z_{2})& \sim &\frac{\frac32 G(z_{2})}{(z_{1}-z_{2})^{2}}  +\frac{\partial_{z_{2}}G(z_{2})}{z_{1}-z_{2}} , \label{eqn:opeN1TG}\\
G(z_{1})G(z_{2})& \sim &\frac{\frac23 c}{(z_{1}-z_{2})^{3}} +\frac{2T(z_{2})}{z_{1}-z_{2}} .  \label{eqn:opeN1GG}
\end{eqnarray}

The first OPE (\ref{eqn:opeN1TT}) is the same as for the bosonic case; thus, it yields the Virasoro algebra, continuous or discrete depending on the choice of the Hamiltonian. The second OPE  (\ref{eqn:opeN1TG}) means that $G(z)$ is a (chiral-) primary field with the dimension $h=\frac32$. Therefore, with the expansion
\begin{equation}
G_{\kappa}\equiv \frac{1}{2\pi i}\oint \kern-.6em \ ^{t}\kern.2em dz g^{\frac32-1}(z)f_{\kappa}(z)G(z),
\end{equation}
the commutation relation
\begin{equation}
[\L_{\kappa}, G_{\kappa'}] = \left(\frac12\kappa -\kappa'\right)G_{\kappa+\kappa'}
\end{equation}
follows directly from Eq. (\ref {eqn:Lphicom}).

The last OPE (\ref{eqn:opeN1GG}) poses an additional issue because both operators are fermionic. For fermionic operators $r^{(j)}(z)$, the time-ordered product is
\begin{equation}
\mathbf{T}\left(r^{(1)}(z_{1}) r^{(2)}(z_{2}) \right)=\left\{\begin{array}{cc} r^{(1)}(z_{1}) r^{(2)}(z_{2})& \hbox{for } t_{1}>t_{2} \\-r^{(2)}(z_{2})r^{(1)}(z_{1})  &  \hbox{for }  t_{1}<t_{2}\end{array}\right. .
\end{equation}
The anti-commutation relations between the integrals of $r^{(j)}(z)$'s along the equal-time contour $R^{(j)}$ can be calculated by ordering $t_{1}$ and $t_{2}$ according to the operator order. The calculation can be represented in the following with the contour shown pictorially:
\begin{eqnarray}
\left\{ R^{(1)}, R^{(2)} \right\}&&=\left\{ \frac{1}{2\pi i} \oint \kern-.6em \ ^{t_{1}}\kern.2em dz_{1}r^{(1)}(z_{1}),  \frac{1}{2\pi i}   \oint \kern-.6em \ ^{t_{2}}\kern.2em dz_{2}r^{(2)}(z_{2}) \right\}  \nonumber \\
&&= \frac{1}{2\pi i}   \oint \kern-.6em \ ^{t_{2}}\kern.2em dz_{2} {\Biggl ( }
\frac{1}{2\pi i}  \int_{\includegraphics[width=3em]{z1z2contour.pdf}} dz_{1} r^{(1)}(z_{1}) r^{(2)}(z_{2}) \nonumber \\
&&\qquad\qquad\quad+\frac{1}{2\pi i}  \int_{\includegraphics[width=3em]{z2z1contour.pdf}} dz_{1}  r^{(2)}(z_{2})r^{(1)}(z_{1})
\Biggr) \\
&&= \frac{1}{2\pi i}   \oint \kern-.6em \ ^{t_{2}}\kern.2em dz_{2} {\Biggl [ }
 \Big( \frac{1}{2\pi i}  \int_{\includegraphics[width=3em]{z1z2contour.pdf}} dz_{1} - \int_{\includegraphics[width=3em]{z2z1contour.pdf}} dz_{1}\Big) \nonumber
 \\&& \qquad \qquad \qquad  \qquad  \qquad \times \mathbf{T}\left(r^{(1)}(z_{1}) r^{(2)}(z_{2}) \right) \Biggl ] \nonumber \\
 &&=\frac{1}{2\pi i}   \oint \kern-.6em \ ^{t_{2}}\kern.2em dz_{2} \int_{\includegraphics[width=1.5em]{z1aroundz2.pdf}} dz_{1} \mathbf{T}\left(r^{(1)}(z_{1}) r^{(2)}(z_{2})  \right) . \nonumber
\end{eqnarray}
Despite the special reference to the point $z=1$, which may infer dipolar quantization, the above description is also valid for ordinary radial quantization.

Because the time-ordered product can be directly related to the OPE, the anti-commutation relation of $G_{\kappa}$'s leads to
\begin{eqnarray}
\left\{ G_{\kappa} , G_{\kappa'} \right\} =&& \frac{1}{2\pi i}   \oint \kern-.6em \ ^{t_{2}}\kern.2em dz_{2} g^{\frac12}(z_{2})f_{\kappa'}(z_{2}) \oint \kern-.6em \ _{z_{2}}\kern.2em \frac{dz_{1}} {2\pi i} g^{\frac12}(z_{1})f_{\kappa}(z_{1})
\nonumber \\ &&  \qquad \qquad \qquad  \times \left(   
\frac{\frac23 c}{(z_{1}-z_{2})^{3}} +\frac{2T(z_{2})}{z_{1}-z_{2}} + \dots 
\right) \ \ .
\end{eqnarray}
While it is  straightforward to see that the second term in the parentheses above simply amounts to
$2\L_{\kappa+\kappa'}$, 
the first term yields the more involved expression:
\begin{equation}
\frac{c}{3}\cdot\kappa^{2}\cdot\frac{1}{2\pi i}   \oint \kern-.6em \ ^{t}\kern.2em dz\frac{f_{\kappa+\kappa'}}{g}
+\frac{c}{3}\cdot\frac{1}{2\pi i}   \oint \kern-.6em \ ^{t}\kern.2em dzf_{\kappa+\kappa'}\left[
\frac12 \frac{d^{2}g}{dz^{2}}-\frac14 (\frac{dg}{dz})^{2}\frac{1}{g}
\right], \label{eqn:GGcenter}
\end{equation}
where the  integration variable  changes to $z$ from $z_{2}$ and the variable dependences of the functions are omitted. The first term in Eq. (\ref{eqn:GGcenter}) is either the Kronecker delta or the Dirac delta function [see Eq. (\ref{eqn:Deltaf})]. For the second term, a separate treatment is necessary. First, we assume $g(z)=z$. Next, $\frac12 \frac{d^{2}g}{dz^{2}}-\frac14 (\frac{dg}{dz})^{2}\frac{1}{g}
$ amounts to $-\frac{1}{4g}$. From Eq.  (\ref{eqn:Deltaf}), this term produces $-\frac14\delta_{\kappa+\kappa', 0}$. On the other hand, if we assume $g(z)=-\frac12(z-1)^{2}$, then this term vanishes. In summary, we obtain
\begin{equation}
\left\{ G_{\kappa} , G_{\kappa'} \right\} =\left\{\begin{array}{cc}2\L_{\kappa+\kappa'}+\frac{c}{3}(\kappa^{2}-\frac14)\delta_{\kappa+\kappa',0} & \hbox{for } g(z)=z \\2\L_{\kappa+\kappa'}+\frac{c}{3}\kappa^{2}\delta(\kappa+\kappa') & \hbox{for } g(z)=-\frac12(z-1)^{2}\end{array}\right. .
\end{equation}

Note that we  see no distinction between Neveau-Schwarz and Ramond fermions for dipolar quantization, $g(z)=-\frac12(z-1)^{2}$. This fact is consistent with the premise that we are dealing with the Riemann sphere which only possesses a unique spin structure.

For $N=2$ SCFT, let us write  the OPEs and the corresponding (anti-)commutation relations without further elaboration because the calculations needed to derive them are rather straightforward. The OPEs we adopt here are
\begin{eqnarray}
T(z_{1})T(z_{2})& \sim &\frac{\frac12 c}{(z_{1}-z_{2})^{4}} + \frac{2T(z_{2})}{(z_{1}-z_{2})^{2}}+\frac{\partial_{z_{2}}T(z_{2})}{z_{1}-z_{2}}, \label{eqn:opeN2TT}\\
T(z_{1})G^{\pm}(z_{2})& \sim &\frac{\frac32 G^{\pm}(z_{2})}{(z_{1}-z_{2})^{2}}  +\frac{\partial_{z_{2}}G^{\pm}(z_{2})}{z_{1}-z_{2}} , \label{eqn:opeN2TG}\\
T(z_{1})J(z_{2})& \sim &\frac{J(z_{2})}{(z_{1}-z_{2})^{2}}  +\frac{\partial_{z_{2}}J(z_{2})}{z_{1}-z_{2}} , \\
J(z_{1})G^{\pm}(z_{2})& \sim &\frac{\pm G^{\pm}(z_{2})}{z_{1}-z_{2}} , \\
G^{+}(z_{1})G^{-}(z_{2})& \sim &\frac{\frac23 c}{(z_{1}-z_{2})^{3}} +\frac{2J(z_{2})}{(z_{1}-z_{2})^{2}}+\frac{2T(z_{2})}{z_{1}-z_{2}} +\frac{\partial_{z_{2}}J(z_{2})}{z_{1}-z_{2}} , \label{eqn:opeN2GG}\\
G^{+}(z_{1})G^{+}(z_{2})& \sim &G^{-}(z_{1})G^{-}(z_{2}) \sim 0 , \\
J(z_{1})J(z_{2})& \sim &\frac{\frac13 c}{(z_{1}-z_{2})^{2}} . 
\end{eqnarray}
We obtain the following (anti-)commutation relations:
\begin{eqnarray}%
\left[\L_{\kappa} , \L_{\kappa'} \right] & = & (\kappa - \kappa')\L_{\kappa+\kappa'}
+\frac{c}{12}\kappa^{3}\delta(\kappa+\kappa') \ , \\
\left[\L_{\kappa} , G^{\pm}_{\kappa'} \right] & = & (\frac12\kappa - \kappa') G^{\pm}_{\kappa+\kappa'} \ , \\
\left[\L_{\kappa} , J_{\kappa'} \right] &=& - \kappa' J_{\kappa+\kappa'} \ , \\
\left[J_{\kappa} , G^{\pm}_{\kappa'} \right] &=& \pm G^{\pm}_{\kappa+\kappa'}  \ , \\
\left\{ G^{+}_{\kappa} , G^{-}_{\kappa'} \right\} &=&2\L_{\kappa+\kappa'}+(\kappa-\kappa')J_{\kappa+\kappa'}+\frac{c}{3}\kappa^{2}\delta(\kappa+\kappa') \ , \\
\left[J_{\kappa}, J_{\kappa'}\right] &=& \frac{c}{3}\kappa \delta(\kappa+\kappa') \ .
\end{eqnarray}
In these equations, we only describe the case of 
dipolar quantization.


\vspace{2em}

\begin{thebibliography}{99}

\bibitem{Heisenberg:1929xj} 
  W.~Heisenberg and W.~Pauli,
  Z.\ Phys.\  {\bf 56}, 1 (1929).

\bibitem{Maldacena:1997re} 
  J.~M.~Maldacena,
  Int.\ J.\ Theor.\ Phys.\  {\bf 38}, 1113 (1999)
  [Adv.\ Theor.\ Math.\ Phys.\  {\bf 2}, 231 (1998)]
  [hep-th/9711200].

\bibitem{Ishibashi:2015jba} 
  N.~Ishibashi and T.~Tada,
  J.\ Phys.\ A {\bf 48}, no. 31, 315402 (2015)
  [arXiv:1504.00138 [hep-th]].


\bibitem{Miyaji:2015fia} 
  M.~Miyaji, T.~Numasawa, N.~Shiba, T.~Takayanagi and K.~Watanabe,
  Phys.\ Rev.\ Lett.\  {\bf 115}, no. 17, 171602 (2015)
  [arXiv:1506.01353 [hep-th]].
  
\bibitem{Verlinde:2015qfa} 
  H.~Verlinde,
  arXiv:1505.05069 [hep-th].

\bibitem{Nakayama:2015mva} 
  Y.~Nakayama and H.~Ooguri,
  JHEP {\bf 1510}, 114 (2015)
  [arXiv:1507.04130 [hep-th]].

\bibitem{Strominger:2013lka} 
  A.~Strominger,
  JHEP {\bf 1407}, 151 (2014)
  [arXiv:1308.0589 [hep-th]].
  
 \bibitem{Belavin:1984vu} 
  A.~A.~Belavin, A.~M.~Polyakov and A.~B.~Zamolodchikov,
  Nucl.\ Phys.\ B {\bf 241}, 333 (1984).
\bibitem{Cardy:1984rp} 
  J.~L.~Cardy,
  J.\ Phys.\ A {\bf 17}, L385 (1984).
  
  
\bibitem{Wybourne:1974}
For example,  see B.~Wybourne,
``Classical Groups for Physicists,''
John Wiley \& Sons, Inc., 1974 [ISBN 0-471-96505-7].






\bibitem{Dirac:1949cp} 
  P.~A.~M.~Dirac,
  Rev.\ Mod.\ Phys.\  {\bf 21}, 392 (1949).
  
\bibitem{Weinberg:1966jm} 
  S.~Weinberg,
  Phys.\ Rev.\  {\bf 150}, 1313 (1966).

\bibitem{Bagchi:2015nca} 
  A.~Bagchi, S.~Chakrabortty and P.~Parekh,
  arXiv:1507.04361 [hep-th].


\bibitem{SSD} A. Gendiar, R. Krcmar and T. Nishino, Prog. Theor. Phys. {\bf 122} (2009) 953; {\it ibid.}  {\bf 123} (2010) 393. 
\bibitem{Proof1} H. Katsura, J. Phys. A: Math. Theor. {\bf 44} (2011) 252001.
\bibitem{Proof2} I. Maruyama, H. Katsura and T. Hikihara, Phys. Rev. {\bf B 84} (2011) 165132.


\bibitem{ShibataHotta2011}
N.~Shibata and C.~Hotta,
Phys.\ Rev.\ B {\bf84} (2011) {115116},
\bibitem{HikiharaNishinio:2011}
T.~Hikihara and T.~Nishino,
Phys.\ Rev.\ B {\bf 83} (2011) 060414.
\bibitem{Gendiaretal2011}
A.~Gendiar, M.~Dani\ifmmode \check{s}\else \v{s}\fi{}ka, Y.~Lee and T.~Nishino,
Phys.\ Rev.\ A {\bf 83} (2011) 052118.
\bibitem{Katsura:2011ss} 
  H.~Katsura,
  J.\ Phys.\ A {\bf 45}, 115003 (2012)


\bibitem{Tada:2014kza} 
  T.~Tada,
  Mod.\ Phys.\ Lett.\ A {\bf 30}, no. 19, 1550092 (2015)
  [arXiv:1404.6343 [hep-th]].
  
  
\bibitem{Tada:2014jps} 
T.~Tada,
{\it Proceedings of the 12th Asia Pacific Physics Conference}
\, JPS Conf. Proc. {\bf 1}, 013003 (2014) .

\bibitem{Holzhey:1994we} 
  C.~Holzhey, F.~Larsen and F.~Wilczek,
  Nucl.\ Phys.\ B {\bf 424}, 443 (1994)
  [hep-th/9403108].
\bibitem{Callan:1994py} 
  C.~G.~Callan, Jr. and F.~Wilczek,
  Phys.\ Lett.\ B {\bf 333}, 55 (1994)
  [hep-th/9401072].
\bibitem{Calabrese:2004eu} 
  P.~Calabrese and J.~L.~Cardy,
  J.\ Stat.\ Mech.\  {\bf 0406}, P06002 (2004)
  [hep-th/0405152].




\bibitem{Garrison:2015lva} 
  J.~R.~Garrison and T.~Grover,
  arXiv:1503.00729 [cond-mat.str-el].

\bibitem{Hotta:2014} 
 C.~Hotta, S.~Nishimoto and N.~Shibata,
 %
 Phys.\ Rev.\ B {\bf 87}, 115128 (2013).




  

\bibitem{Fubini:1972mf} 
  S.~Fubini, A.~J.~Hanson and R.~Jackiw,
  Phys.\ Rev.\ D {\bf 7}, 1732 (1973).
  



\bibitem{Luscher:1974ez} 
  M.~Luscher and G.~Mack,
  Commun.\ Math.\ Phys.\  {\bf 41}, 203 (1975).
  doi:10.1007/BF01608988
  
\bibitem{Rychkov:2016iqz} 
  S.~Rychkov,
``EPFL Lectures on Conformal Field Theory in $D\geq 3$ Dimensions,''
  arXiv:1601.05000 [hep-th].

\bibitem{Polchinski:1998rq} 
  J.~Polchinski,
``String theory. Vol. 1: An introduction to the bosonic string,''
Cambridge University Press 1998  [ISBN 0-521-63303-6].

\bibitem{Aoki:1998vn} 
  H.~Aoki, S.~Iso, H.~Kawai, Y.~Kitazawa and T.~Tada,
  Prog.\ Theor.\ Phys.\  {\bf 99}, 713 (1998)
  [hep-th/9802085].

  
      
\bibitem{Zamolodchikov:1986gt}
  A.~B.~Zamolodchikov,
  JETP Lett.\  {\bf 43} (1986) 730
   [Pisma Zh.\ Eksp.\ Teor.\ Fiz.\  {\bf 43} (1986) 565].
\bibitem{Polchinski:1987dy} 
  J.~Polchinski,
  Nucl.\ Phys.\ B {\bf 303}, 226 (1988).
 

\bibitem{Okunishi:2015dfa} 
  K.~Okunishi and H.~Katsura,
  J.\ Phys.\ A {\bf 48}, no. 44, 445208 (2015)
  [arXiv:1505.07904 [cond-mat.stat-mech]].
  
\bibitem{Ramond:2010zz} 
  P.~Ramond,
 ``Group theory: A physicist's survey,''
  Cambridge University Press 2010 [ISBN 0-521-89603-7].

\bibitem{Kiermaier:2007jg} 
  M.~Kiermaier, A.~Sen and B.~Zwiebach,
  JHEP {\bf 0803}, 050 (2008)
  [arXiv:0712.0627 [hep-th]].

\bibitem{Takahashi:2003xe} 
  T.~Takahashi and S.~Zeze,
  Prog.\ Theor.\ Phys.\  {\bf 110}, 159 (2003)
  [hep-th/0304261].
  
  
\bibitem{Takahashi:2011zzb} 
  T.~Takahashi,
  Prog.\ Theor.\ Phys.\ Suppl.\  {\bf 188}, 163 (2011).

\bibitem{Maldacena:2005hi} 
  J.~M.~Maldacena,
   JHEP {\bf 0509}, 078 (2005)
  [Int.\ J.\ Geom.\ Meth.\ Mod.\ Phys.\  {\bf 3}, 1 (2006)]
  [hep-th/0503112].
\bibitem{Seiberg:2005nk} 
  N.~Seiberg,
  JHEP {\bf 0601}, 057 (2006)
  [hep-th/0511220].
 
\bibitem{Kostov:2006dy} 
  I.~Kostov,
  JHEP {\bf 0701}, 074 (2007)
  [hep-th/0610084].



\bibitem{GradshteynRyzhik}
A.~Jeffrey and D.~Zwillinger (eds.),
``Gradshteyn and Ryzhik's Table of Integrals, Series, and Products,''
Academic Press 2007 [ISBN 0-123-73637-4].

\end{thebibliography}
\end{document}